%% file: main.tex
\documentclass[10pt,twocolumn,letterpaper]{article}

\usepackage[pagenumbers]{cvpr} % To force page numbers, e.g. for an arXiv version

\usepackage{graphicx}
\usepackage{amsmath}
\usepackage{amssymb}
\usepackage{float}
\usepackage{booktabs}
\usepackage[dvipsnames]{xcolor}
\usepackage[pdftex,outline]{contour}
\usepackage{xcolor}
\usepackage{overpic}

\input{macros}

\usepackage[pagebackref,breaklinks,colorlinks]{hyperref}

\usepackage[capitalize]{cleveref}
\crefname{section}{Sec.}{Secs.}
\Crefname{section}{Section}{Sections}
\Crefname{table}{Table}{Tables}
\crefname{table}{Tab.}{Tabs.}

\def\cvprPaperID{6045} % *** Enter the CVPR Paper ID here
\def\confName{CVPR}
\def\confYear{2023}

\begin{document}

%%%%%%%%% TITLE
% \title{3D Highlighter: Selecting 3D Shape Regions via Text Descriptions}
\title{3D Highlighter: Localizing Regions on 3D Shapes via Text Descriptions}

%%%%%%%%% Authors
\input{00_authors.tex}

%%%%%%%%% TEASER
\twocolumn[{%
\renewcommand\twocolumn[1][]{#1}%
\maketitle
\input{figures/teaser}
}]

%%%%%%%%% ABSTRACT
\input{00_abstract.tex}

%%%%%%%%% TEXT BODY
\input{01_introduction.tex}
\input{02_related_work.tex}
\input{03_method.tex}
\input{04_experiments.tex}

\input{05_conclusions.tex}

%%%%%%%%% REFERENCES
{\small
\bibliographystyle{ieee_fullname.bst}
\bibliography{references.bib}
}

%%%%%%%%% Supplementary Material
% \clearpage
\input{07_supplementary}

\end{document}

%% file: macros.tex
\newif\ifdraft
\draftfalse

\DeclareMathOperator*{\argmin}{\arg\!\min}

\ifdraft
\newcommand{\rhc}[1]{{\color{blue}[\textbf{Rana:} #1]}}
\newcommand{\dale}[1]{{\color{red}[\textbf{Dale:} #1]}}
\newcommand{\itai}[1]{{\color{green}[\textbf{Itai:} #1]}}

\else
\newcommand{\rhc}[1]{}
\newcommand{\dale}[1]{}
\newcommand{\itai}[1]{}

\fi

\newcommand{\ourmethod}{3D Highlighter}

%% file: 00_authors.tex
\author{
Dale Decatur \\
University of Chicago \\
%Institution address \\
{\tt\small ddecatur@uchicago.edu}
\and
Itai Lang\\
University of Chicago \\
%Institution address \\
{\tt\small itailang@uchicago.edu}
\and
Rana Hanocka \\
University of Chicago \\
%Institution address \\
{\tt\small ranahanocka@uchicago.edu} \\ \\
}

%% file: figures/teaser.tex
\begin{center}
    \centering
    \newcommand{\pl}{-2}
    \begin{overpic}[width=\textwidth, trim=0 0 0 40]{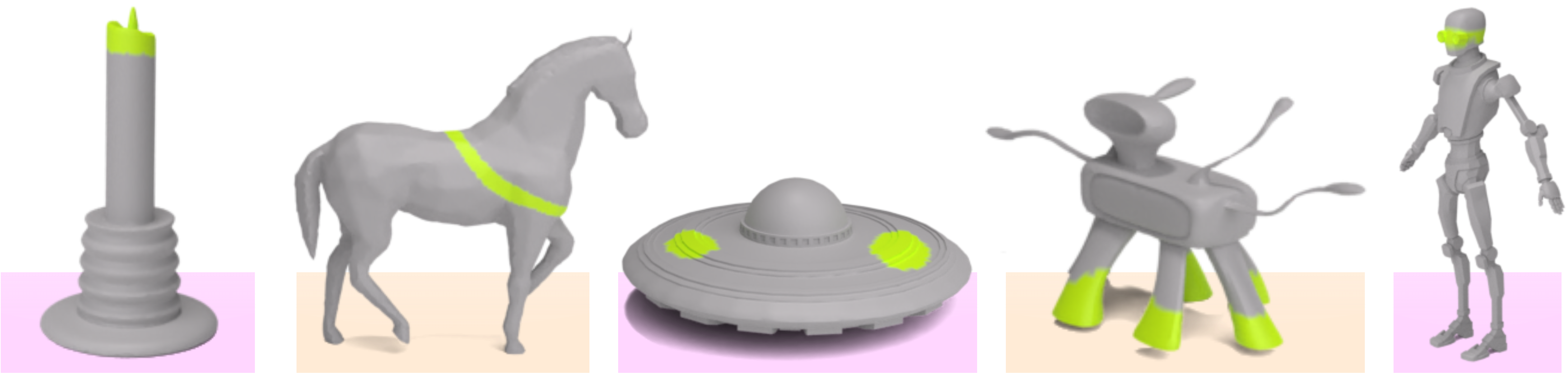}
    \put(7,  \pl){\textcolor{black}{Hat}}
    \put(24,  \pl){\textcolor{black}{Necklace}}
    \put(46,  \pl){\textcolor{black}{Headlights}}
    \put(73,  \pl){\textcolor{black}{Shoes}}
    \put(90,  \pl){\textcolor{black}{Eyeglasses}}
    \end{overpic}
    \vspace{-2mm}
    \captionof{figure}{\ourmethod{} localizes semantic regions on a shape using text as input. Our technique reasons about \textit{where} to place seemingly unrelated concepts in semantically meaningful locations on the 3D shape, such as a `necklace' on a horse or `shoes' on an alien.}
    \label{fig:teaser}
\end{center}

%% file: 00_abstract.tex
\begin{abstract}
We present 3D Highlighter, a technique for localizing semantic regions on a mesh using text as input. A key feature of our system is the ability to interpret ``out-of-domain" localizations. Our system demonstrates the ability to reason about where to place non-obviously related concepts on an input 3D shape, such as adding clothing to a bare 3D animal model. Our method contextualizes the text description using a neural field and colors the corresponding region of the shape using a probability-weighted blend. Our neural optimization is guided by a pre-trained CLIP encoder, which bypasses the need for any 3D datasets or 3D annotations. Thus, 3D Highlighter is highly flexible, general, and capable of producing localizations on a myriad of input shapes. Our code is publicly available at \url{https://github.com/threedle/3DHighlighter}.
\end{abstract}

%% file: 01_introduction.tex
\vspace{-3mm}
\section{Introduction}
\label{sec:intro}
%%%%%%%%%%%%%%%%%%%%%%%%%%%%%%%%%
%% the motivation to our work %%%
%%%%%%%%%%%%%%%%%%%%%%%%%%%%%%%%%

Semantic localization of regions on 3D meshes is an important problem in computer graphics and vision with broad applications. One such application is the incorporation of semantic information into the 3D modeling process. 
A particularly challenging aspect of this task emerges when 3D geometric signals are insufficient for performing segmentation, \eg where to add a shirt to a bare 3D human model.

%%%%%%%%%%%%%%%%%%%%%%%%%%%%%%%%%%%%%%%%%%%%%
%% high level overview of our method does %%%
%%%%%%%%%%%%%%%%%%%%%%%%%%%%%%%%%%%%%%%%%%%%%
We propose \textit{\ourmethod{}}, a method for automatically localizing \textit{fine-grained} semantic regions on a shape based on only a text description. Our system contextualizes the text prompt and \textit{highlights} the corresponding shape region using the network-predicted probabilities. Using \textit{only} text, users are able to semantically identify regions on a shape.
Our system takes meshes as input, making it compatible with 3D modeling workflows and tools.

%%% Gallery Figure %%%
\input{figures/gallery}

%%%%%%%%%%%%%%%%%%%%%%%%%%%%%%%%%%
%%% demonstrated understanding %%%
%%%%%%%%%%%%%%%%%%%%%%%%%%%%%%%%%%
This highlighting task requires both object-level and part-level understanding. \ourmethod{} demonstrates the ability to reason about \textit{where} to place seemingly unrelated concepts on the 3D shape, such as a hat on a lamp (\cref{fig:teaser}). Our system localizes attributes that are \textit{geometrically absent} from a shape, which we refer to as \emph{hallucinated highlighting}. Understanding a part's global shape context is challenging even when relying on salient geometric features~\cite{kaick2014shape, kaichun2018partnet}, let alone without them.

%%%%%%%%%%%%%%%%%%%%
%%% how we do it %%%
%%%%%%%%%%%%%%%%%%%%
We optimize the weights of a neural network to produce probabilities that are used to color a given 3D shape in accordance with the specified text. We leverage a pre-trained vision-language model (CLIP \cite{CLIP}) to guide the neural optimization towards the text-specified region. This neural optimization formulation is flexible, bypassing the need for any 3D datasets, 3D annotations, or 3D pre-training. Our system is not bound to a specific set of classes, and, as shown in \cref{fig:gallery}, is not limited to object parts defined by salient geometric features. 

We encode the part selection as a \textit{neural field}~\cite{brownneuralfields} over the mesh surface. Our network learns to map each point on the surface to a probability of belonging to the text-specified region. We translate the inferred probabilities to a visual attribute on the mesh surface, which can be rendered and visually understood. The network-predicted probabilities act as a soft-selection operator which \textit{blends} the highlighter color onto the mesh. The network weights are updated by encouraging the CLIP~\cite{CLIP} embedding of the 2D renders of the highlighted mesh to adhere to the specified text. As a result, the network implicitly learns to segment the object to adhere to the text prompt.

% multi-target figure
\input{figures/multi_target}

%%%%%%%%%%%%%%%%%%%%%%%%%
%%% Key to our method %%%
%%%%%%%%%%%%%%%%%%%%%%%%%
We make several design choices that are key to the success of \ourmethod{}. 
Our network does not directly color the mesh. Rather, we predict a \textit{probability of being inside the text-specified highlight}, which is used to blend colors on the mesh. The network is initialized such that points have roughly a 50\% probability of being highlighted, resulting in a mesh with albedo halfway between the highlight and background color. During optimization, the relative blend weight of the highlight color directly corresponds to the highlight probability. This blending enables the network to naturally and smoothly increase or decrease the segmentation probability in accordance with the text specification of the target region.

%%%%%%%%%%%%%%%%
%%% summary  %%%
%%%%%%%%%%%%%%%%
In summary, we present a method for localizing semantic regions on 3D shapes. The localization is specified by a textual description, which is intuitive, flexible, and not limited to a specific training dataset. We demonstrate applications of our method to shape editing and stylization. Furthermore, our field formulation enables the \ourmethod{} to work with different mesh resolutions and triangulations. A key feature of our system is the ability to interpret out-of-domain localizations. For example, \ourmethod{} is able to figure out where to place a `hat' on a candle as seen in \cref{fig:teaser}, demonstrating the ability to reason about \textit{where} to place seemingly unrelated concepts on the 3D shape.

%% file: figures/gallery.tex
\begin{figure*}[t!]
    \centering
    \newcommand{\rowone}{54}
    \newcommand{\rowtwo}{36}
    \newcommand{\rowthree}{17}
    \newcommand{\rowfour}{-2}
    \begin{overpic}[width=\textwidth]{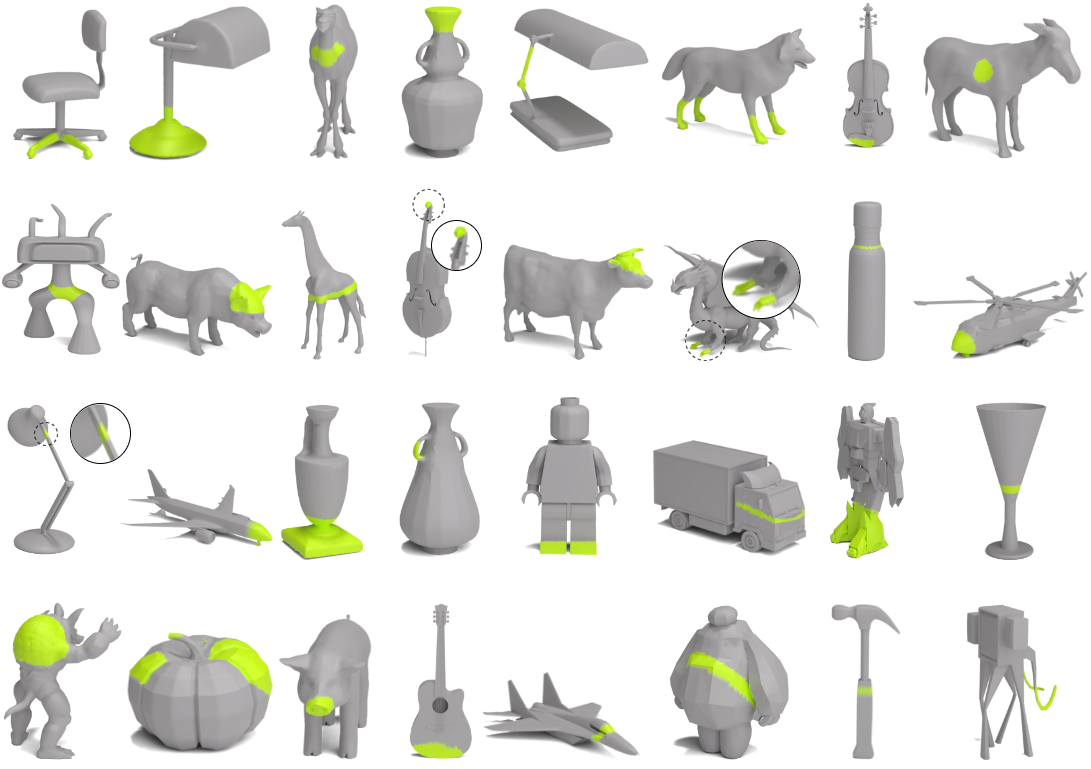}
    \put(3,  \rowone){\textcolor{black}{Shoes}}
    \put(13,  \rowone){\textcolor{black}{Wheels}}
    \put(27,  \rowone){\textcolor{black}{Bow Tie}}
    \put(39,  \rowone){\textcolor{black}{Hair}}
    \put(48,  \rowone){\textcolor{black}{Necklace}}
    \put(64,  \rowone){\textcolor{black}{Shoes}}
    \put(77,  \rowone){\textcolor{black}{Shoes}}
    \put(90,  \rowone){\textcolor{black}{Heart}}
    \put(3,  \rowtwo){\textcolor{black}{Belt}}
    \put(17,  \rowtwo){\textcolor{black}{Hat}}
    \put(29,  \rowtwo){\textcolor{black}{Belt}}
    \put(38,  \rowtwo){\textcolor{black}{Hat}}
    \put(51,  \rowtwo){\textcolor{black}{Hat}}
    \put(65,  \rowtwo){\textcolor{black}{Gloves}}
    \put(76,  \rowtwo){\textcolor{black}{Necklace}}
    \put(90,  \rowtwo){\textcolor{black}{Mask}}
    \put(1,  \rowthree){\textcolor{black}{Necklace}}
    \put(16,  \rowthree){\textcolor{black}{Snout}}
    \put(27,  \rowthree){\textcolor{black}{Shoes}}
    \put(38,  \rowthree){\textcolor{black}{Arm}}
    \put(50,  \rowthree){\textcolor{black}{Shoes}}
    \put(64,  \rowthree){\textcolor{black}{Mouth}}
    \put(77,  \rowthree){\textcolor{black}{Floor}}
    \put(90,  \rowthree){\textcolor{black}{Scarf}}
    \put(3,  \rowfour){\textcolor{black}{Roof}}
    \put(16,  \rowfour){\textcolor{black}{Wings}}
    \put(24,  \rowfour){\textcolor{black}{Car Headlights}}
    \put(38,  \rowfour){\textcolor{black}{Shoes}}
    \put(50,  \rowfour){\textcolor{black}{Glasses}}
    \put(65,  \rowfour){\textcolor{black}{Belt}}
    \put(77,  \rowfour){\textcolor{black}{Collar}}
    \put(90,  \rowfour){\textcolor{black}{Braids}}
    \end{overpic}
    \vspace{-2mm}
    \caption{\textbf{Hallucinated part highlighting}. Our system is able to reason about \textit{where} to highlight a \textit{geometrically-absent} region on shapes. The resulting localizations demonstrate global understanding and localized part-awareness.}
    \label{fig:gallery}
\end{figure*}
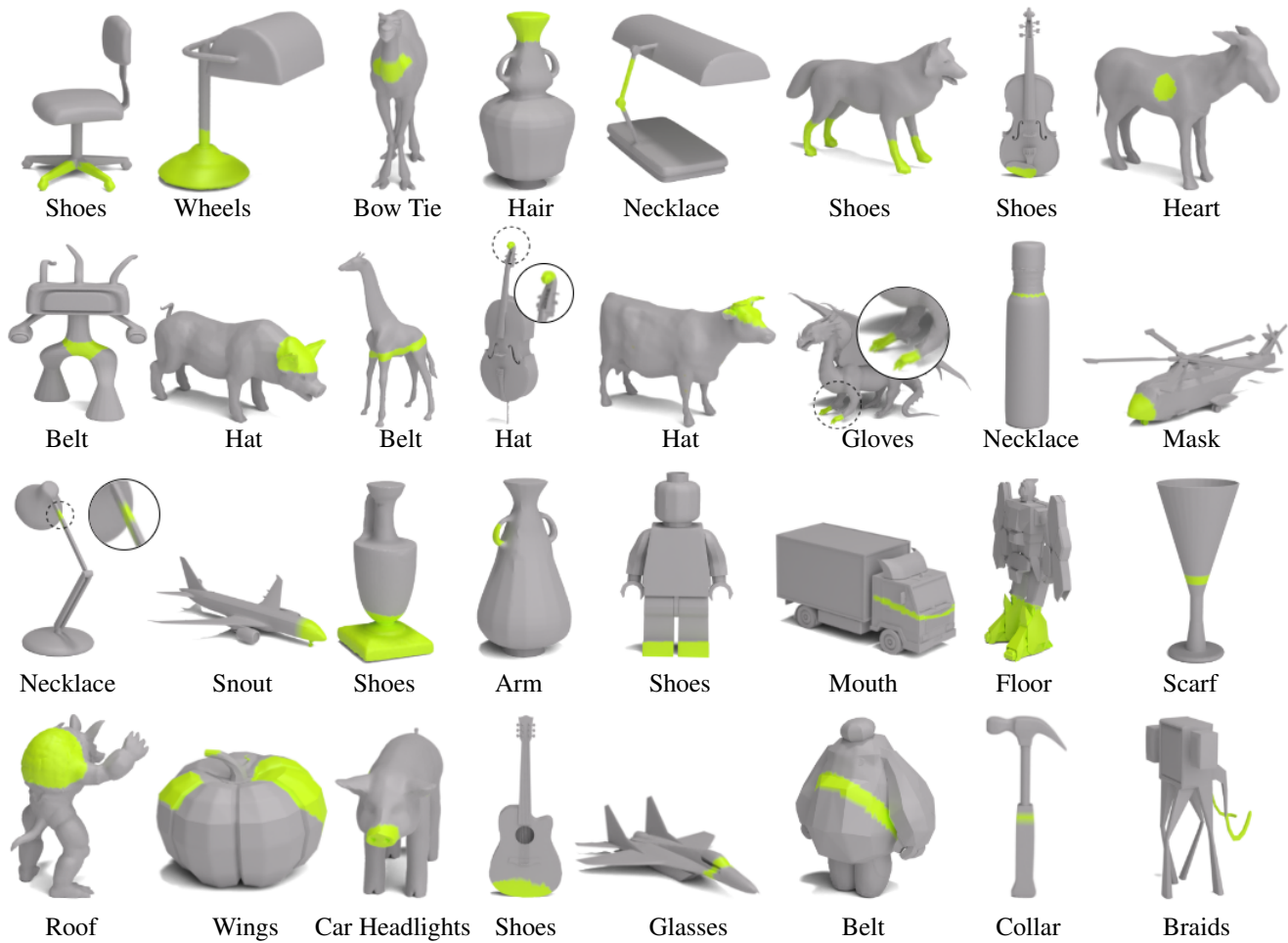

%% file: figures/multi_target.tex
\begin{figure}[t!]
    \centering
    \newcommand{\one}{75}
    \newcommand{\two}{52}
    \newcommand{\three}{28}
    \newcommand{\four}{-2}
    \begin{overpic}[width=\columnwidth]{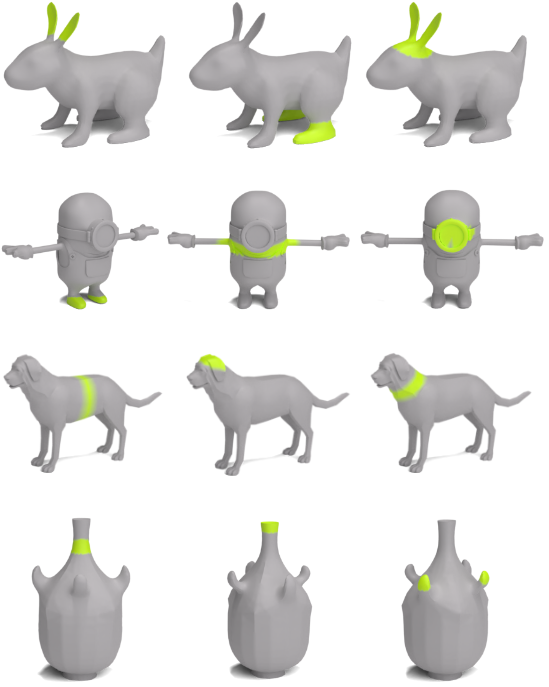}
    \put(5,  \one){Headphones}
    \put(37,  \one){Shoes}
    \put(63,  \one){Hat}
    \put(8,  \two){Shoes}
    \put(31,  \two){Necklace}
    \put(60,  \two){Glasses}
    \put(9,  \three){Belt}
    \put(36,  \three){Hat}
    \put(59,  \three){Necklace}
    \put(5,  \four){Necklace}
    \put(36,  \four){Roof}
    \put(61,  \four){Arms}
    \end{overpic}
    \caption{Our method is able to highlight different parts on the same object. For target selections that correspond to distinct regions, \ourmethod{} produces selections that are semantically meaningful and spatially separated without signal from underlying geometry.}
    \label{fig:multi_target}
\end{figure}
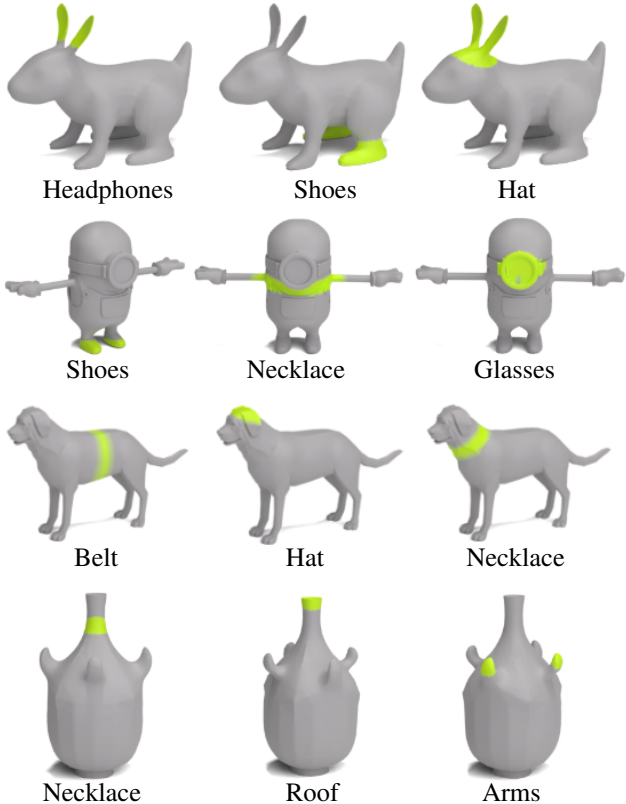

%% file: 02_related_work.tex
\section{Related Work}
\label{sec:rw}

\smallskip
\noindent \textbf{Geometry-driven segmentation.} Traditional works in geometry processing use low-level geometric features (such as surface area, curvature, or geodesic distance) in order to infer high-level semantic attributes for segmenting shapes~\cite{shamir2008survey}. In particular, decomposing shapes into smaller parts or segments often corresponds with physical 3D semantic parts~\cite{hoffman1984parts, shamir2008survey}. One approach is to partition shapes based on convexity, or an approximation of convexity~\cite{lien2007approximate, asafi2013weak}. The medial axis carries topological information, which may also be used as a guideline for segmentation~\cite{skeleton_intrinsic, cornea2007curve, dey2004approximating, shamir2008survey}.

The underlying assumption in these works is that processing the local geometry can be used to understand the semantics for segmentation. By contrast, a key aspect of our work is the ability to perform hallucinated highlights: segmentations that can not necessarily be inferred by geometry alone. See example highlights in \cref{fig:gallery} (e.g., localizing a heart on a goat).

\smallskip
\noindent \textbf{Data-driven segmentation.} In the deep learning era, the 3D part segmentation task has been widely tackled by neural network models~\cite{yi2017syncspeccnn, hanocka2019meshcnn, lahav2020meshwalker, milano2020primal, sharp2022diffusionnet, subdivnet}. Training such a model is typically done in a fully-supervised manner on a large dataset of shapes annotated with a given set of part classes. For example, MeshCNN~\cite{hanocka2019meshcnn} was trained on a human-body segmentation dataset~\cite{maron2017convolutional} for learning semantic part segmentation. To alleviate the need for 3D annotations, unsupervised learning schemes utilize large collections of unlabelled data~\cite{chen2019bae, deng2020cvxnet, zhu2020adacoseg, canonical_capsules, hong2022threedcg}. For example, Hong \etal~\cite{hong2022threedcg} inferred part-segmentation through question answering on rendered images from PartNet~\cite{yu2019partnet}. 

In contrast to existing deep learning approaches for shape segmentation, we do not rely on any 3D dataset, nor are we bounded to a specific shape category or set of parts. Instead, we specify the desired localization using text and a pre-trained CLIP model which encompasses rich semantic object understanding. Thus, our \ourmethod{} is capable of localizing various semantic regions on a wide variety of 3D shapes.

\input{figures/edit}

\input{figures/overview}

\smallskip
\noindent \textbf{Text-guidance.}
Recent works have leveraged pre-trained vision-language embedding spaces, such as CLIP~\cite{CLIP}, for analysis, synthesis, and editing. Some techniques leverage pre-trained image encoders for achieving semantic segmentation in images and neural radiance fields~\cite{lseg, text2live, kobayashi2022distilledfeaturefields}. Such techniques are capable of segmenting \textit{entire} objects within a scene based on text, \eg, a chair inside a room. However, they may struggle to segment \textit{parts} within an object; \eg, failing to distinguish a window (part) from a house (object)~\cite{lseg}.

Our work is inspired by the emergent analysis in text-driven synthesis techniques for 3D data~\cite{text2mesh, dreamfields, clipmesh, dreamfusion}. Specifically, Text2Mesh~\cite{text2mesh} devised a framework for text-driven stylization of 3D meshes, observing that the resulting textures consider part-aware semantics. Yet, since Text2Mesh directly synthesizes stylizations, there is no obvious way to extract any underlying semantic analysis. To address this, we opt to use a highlighter color only as a means for visualizing the network-predicted segmentations.

%% file: figures/edit.tex
\begin{figure}[b]
    \centering
    \newcommand{\rowone}{-3}
    \begin{overpic}[width=\columnwidth]{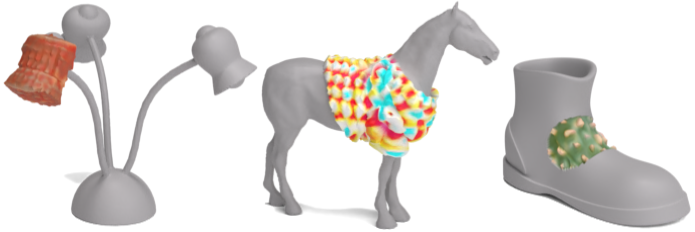}
    \put(10,  \rowone){Flower}
    \put(44,  \rowone){Poncho}
    \put(78,  \rowone){Laces}
    \end{overpic}
    \captionof{figure}{\textbf{Localized editing.} We incorporate textures and displacements to a region highlighted with \ourmethod{}. Used styles: Brick (left), Colorful Crochet (middle), Cactus (right).}
    \label{fig:edit}
\end{figure}

%% file: figures/overview.tex
\definecolor{highlighter}{rgb}{0.7, 0.85, 0.0}
\definecolor{input}{rgb}{0.5, 0.85, 0.6}
\begin{figure*}[t!]
    \centering
    \newcommand{\rendcolor}{\color[rgb]{0.5,0.5,0.9}}
    \includegraphics[width=\textwidth]{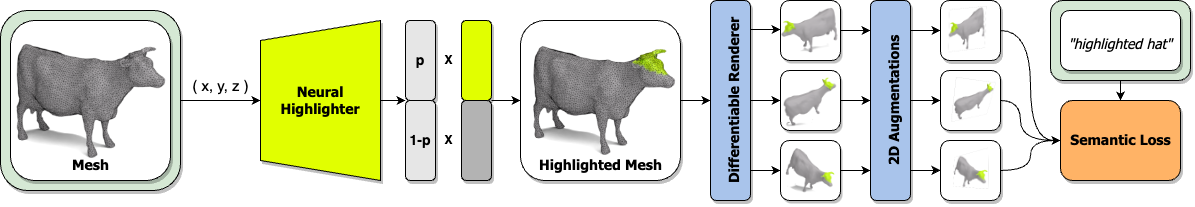}
    \caption{\textbf{Overview of \ourmethod.} The \textcolor{highlighter}{\textbf{Neural Highlighter}} maps each point on the \textcolor{input}{\textbf{input mesh}} to a \textit{probability}. The mesh is colored using a probability-weighted blend and then {\rendcolor \textbf{rendered}} from multiple views. The neural highlighter weights are guided by the similarity between the  \textcolor{Dandelion}{\textbf{CLIP embeddings}} of the 2D augmented images and the \textcolor{input}{\textbf{input text}}.}
    \label{fig:overview}
\end{figure*}

%% file: 03_method.tex
\section{Method}

\label{sec:method}

An illustration of our method is shown in Fig.~\ref{fig:overview}. The inputs to our system are a mesh $M$, represented by vertices $V \in \mathbb{R}^{n \times 3}$ and faces $F \in \{1,...,n\}^{m \times 3}$, and a text description $T$. Our neural network, referred to as \emph{neural highlighter}, is optimized to map vertex positions $v \in V$ to a probability $p$ of belonging to the text-specified region. Each vertex on the mesh is colored according to a probability-weighted blend between the highlighter color and a gray background color. The resulting highlighted mesh $M'$ is rendered from multiple views, and we apply 2D augmentations to obtain a set of images. We supervise the network optimization by comparing the CLIP-embedded images to the CLIP embedding of the desired text.

\input{figures/viewpoints}
\subsection{Neural Highlighter}
\label{subsec:neuralhighlighter}
Our neural highlighter is a neural field~\cite{brownneuralfields} mapping coordinates $\mathbf x \in \mathbb{R}^3$ to $p \in [0, 1]$, where $p$ is the probability that $\mathbf x$ belongs to the text-specified region. The neural highlighter is represented as a multi-layer perceptron (MLP) $\mathcal{F}_\theta$ that takes an input vertex $v$ in the form of a 3D coordinate $\mathbf x_v=(x,y,z)$ and predicts a highlight probability $p_v$, $\mathcal{F}_\theta(\mathbf x_v) = p_v$. This formulation allows us to query the neural field to obtain meaningful highlight probabilities for any 3D point on (or near) the mesh surface. Thus, once optimized, the network weights conveniently transfer the localization to different meshings of the same object without requiring further optimization (\cref{fig:invariance}).

Representing our neural highlighter as an MLP produces contiguous localizations and reduces artifacts. MLPs have been shown to exhibit a spectral bias towards smooth solutions~\cite{spectralbias}, especially on low-dimensional inputs such as 3D coordinates~\cite{tancik2020fourfeat}. The bias towards low-frequency outputs encourages our \ourmethod{} to predict contiguous localizations with sharp boundaries and discourages noisy highlights (\cref{fig:ablation}). For this reason, our approach does not utilize positional encoding. See supplemental material for additional details.

\subsection{Mesh Color Blending}
We leverage the per-point highlight probability to color the mesh in a continuous, differentiable manner, generating semantically meaningful renders for CLIP supervision. We use a probability-weighted blend, where each vertex color $C_v$ is a linear combination of the highlight color $H$ and gray color $G$ weighted by the network-predicted highlight probability $C_v = p_v \cdot H + (1 - p_v) \cdot G$.

At the start of the optimization process, all vertex probabilities are initialized near $0.5$ and thus the entire mesh is half-highlighted. As the optimization progresses, vertices smoothly transition towards gray or highlighter color (based on the network predictions) such that vertices predicted to be highlighted adhere to the text-specified region.
This formulation translates each step of the optimization to a colored mesh that is semantically meaningful to CLIP.
Our method provides continuous gradients, in contrast to coloring vertices according to the argmax of the highlight probability. Our blending scheme results in a smoother optimization landscape and reduces highlight artifacts (\cref{fig:ablation}).

This formulation is also important for downstream applications that wish to use the localizations, \eg editing and stylization. Predicting per-point highlight probabilities provides an explicit representation of the highlight region on the mesh surface. An alternative approach, optimizing the surface color directly, would only provide a visual result without explicit information about which vertices belong to the localization.

\subsection{Unsupervised Guidance}
\label{subsec:guidance}
We guide our neural optimization using the joint vision-language embedding space of CLIP~\cite{CLIP}. We formulate the desired highlight by describing the association between the input mesh [object] and target localization [region]. Specifically, we design our target text $T$ to be: ``a gray [object] with highlighted [region]." 
We render the highlighted geometry from multiple views using differentiable rendering~\cite{chen2019learning}. At each optimization step, we randomly sample $n$ views from a Gaussian distribution centered around a primary view. This ensures that the underlying object is recognizable in the majority of views shown to CLIP.

In a preliminary viewpoint prediction stage, we render $360^{\circ}$ views of the mesh and measure the CLIP similarity to the target text prompt. We select the primary view to be the render with the highest CLIP similarity. We found that there exist many possible viewpoints which produce desirable highlighter results (see \cref{fig:viewpoints}). More details about how the primary view is selected can be found in the supplemental material.

For each view $\psi$, we render a 2D image $I_\psi$ and apply a random perspective 2D augmentation $\phi$, as done in previous works~\cite{text2mesh, clipdraw}. We then encode each of the augmented images into the CLIP embedding space (in $\mathbb{R}^{768}$) using CLIP's image encoder, denoted as $E_I$. Our final aggregate image representation $\mathsf e_I$ is the average CLIP encoding over all views:
\begin{equation}
    \mathsf e_I = \frac{1}{n}\sum_{\psi} E_I(\phi(I_\psi)) \in \mathbb{R}^{768}.
\end{equation}
Similarly, we encode the target selection text $T$ with CLIP's text encoder $E_T$ to get the encoded target representation $\mathsf e_T = E_T(T) \in \mathbb{R}^{768}$. Our loss $\mathcal{L}$ for optimizing the neural highlighter parameters $\theta$ is formulated as the negative cosine similarity between the aggregate image embedding and the text embedding:
\begin{equation}
    \argmin_\theta \mathcal{L(\theta)} = - \frac{\mathsf e_I \cdot \mathsf e_T}{|\mathsf e_I| \cdot |\mathsf e_T|}.
\end{equation}
\noindent When the loss is minimized, the CLIP embedding of the rendered highlighted mesh becomes similar to the target text embedding. Thus, the localized region will reflect the target text region.

%% file: figures/viewpoints.tex
\begin{figure}[b]
    \centering
    \newcommand{\dblue}{\color[rgb]{0.56,0.82,0.92}}
    \newcommand{\rowone}{-3}
    \begin{overpic}[width=\columnwidth]{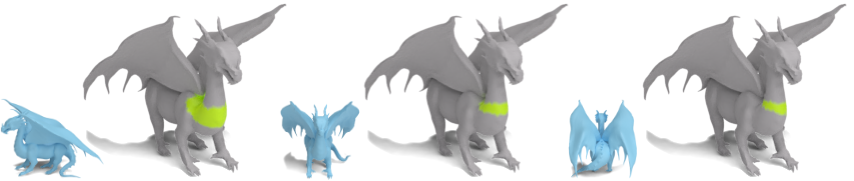}
    \put(13,  \rowone){$0^{\circ}$}
    \put(44,  \rowone){$90^{\circ}$}
    \put(75,  \rowone){$-90^{\circ}$}
    \end{overpic}
    \captionof{figure}{\textbf{Viewpoint robustness.} Our system produces consistent results even when using different \textit{primary} viewpoints. Results for {\dblue \textbf{three different primary viewpoints}} for the target text `necklace'.}
    \label{fig:viewpoints}
\end{figure}

%% file: 04_experiments.tex
\section{Experiments}
\label{sec:experiments}
In this section we examine various capabilities of \ourmethod{}. First, we demonstrate the fidelity of our highlighter localization in \cref{subsec:fidelity}, including qualitative and quantitative evaluations.
As far as we can ascertain, our method is the first technique to perform text-driven localization on 3D shapes without pre-training on 3D data. Thus, we adapt an existing language-guided segmentation technique for 2D images to serve as a baseline~\cite{lseg}. Moreover, we demonstrate the robustness of \ourmethod{} in \cref{subsec:generality}. Then we explore several applications of our method in \cref{subsec:applications}, such as selective editing, localized manipulation, and segmentation. Finally, in \cref{subsec:components} we evaluate the influence of key components of \ourmethod{} and discuss its limitations in \cref{subsec:limitations}.

We apply our method to a large variety of meshes from different sources: COSEG~\cite{coseg_2011}, Turbo Squid~\cite{turbosquid}, Thingi10K~\cite{Thingi10K}, Toys4k~\cite{Toys4k}, ModelNet~\cite{modelnet}, and ShapeNet~\cite{chang2015shapenet}. \ourmethod{} does not impose any restrictions on the mesh quality; many of the meshes used contain artifacts, such as elements that are non-manifold, unoriented, and contain boundaries or self-intersections. Our PyTorch~\cite{pytorch} implementation optimization takes around 5 minutes to run on an Nvidia A40 GPU. In our experiments, we used CLIP ViT-L/14 at $224 \times 224$ resolution.

\input{figures/ablation}
%%%%%%%%%%%%%%%%
%%% Fidelity %%%
%%%%%%%%%%%%%%%%
\subsection{Generality and Fidelity of 3D Highlighter}
\label{subsec:fidelity}

\noindent \textbf{Highlight generality.}
\ourmethod{} is not restricted to any particular category for either the input mesh or the text-specified localization, since it does not rely on a 3D dataset or 3D pre-training. In \cref{fig:gallery}, we see our method achieves accurate localization for a diverse collection of meshes from various domains such as humanoids, animals, and manufactured objects. \ourmethod{} is capable of localizing a wide variety of diverse attributes even when the context of these target attributes is entirely unrelated to the input mesh. Moreover, \ourmethod{} demonstrates that it can perform \textit{hallucinated highlighting}, where it selects regions on meshes with no underlying geometric signal (such as a bow tie on a camel or a hat on a pig).

\smallskip
\noindent \textbf{Highlight specificity.} In \cref{fig:multi_target}, we observe that semantic differences are reflected in the network-predicted highlight. \ourmethod{} is able to successfully localize different text-specified regions on the same mesh. Our framework demonstrates the nuanced understanding required to disambiguate different target regions, such as headphones and hat on the rabbit. Finally, the ability to identify many different regions on a single mesh allows users intuitive, comprehensive, and fine-grained control over part localization. 

\input{tables/clip_metrics}

\smallskip
\noindent \textbf{Quantitative evaluation.}
\ourmethod{} is the first system to select semantic regions on 3D shapes using text guidance, without any 3D datasets. Since there are no quantitative benchmarks to evaluate the quality of our highlights, we built an evaluation strategy inspired by Dream Fields~\cite{dreamfields, park2021benchmark}. In our work, we build a specialized CLIP R-Precision metric based on text-retrieval with text prompts created for different localizations of semantic regions.

% how we did it (the comparison in 2D)
Moreover, since there are no existing approaches for text-based segmentation in 3D, we create two baselines by extending two different 2D image-based approaches. The first baseline extends LSeg~\cite{lseg} which directly predicts segmentation in 2D, while the second baseline extends Text2LIVE~\cite{text2live} which infers an edit mask for 2D image manipulation. To evaluate these baselines, we render a bare mesh from a view where the target localization region is clearly visible. We extract the 2D segmentation produced by the image baselines and use it to color the rendered image. Then we evaluate both baselines and our 3D Highlighter result rendered from the same view using the specialized CLIP R-Precision.

% what we see / summarizing the table
We run this evaluation on the ViT-L/14 and ViT-B/16 CLIP models and observe that our method outperforms both baselines (see \cref{table:clip_metric}). LSeg is built for text-driven semantic segmentation and excels at segmenting entire objects within a scene. However, LSeg struggles to identify parts within a single object, leading to subpar performance on our highlighting task. Text2LIVE was not explicitly built for segmentation, however it does rely on inferring a continuously-valued edit mask (\ie a soft-segmentation) when performing localized image editing. The edit mask is designed to produce high-quality image manipulations; however, it is not directly suitable for identifying the sharp segmentation boundaries required for our highlighting task. Details and qualitative examples are provided in the supplemental material.

\input{figures/swap}
%%%%%%%%%%%%%%%%%%
%%% Generality %%%
%%%%%%%%%%%%%%%%%%
\subsection{Robustness of \ourmethod{}}
\label{subsec:generality}
\input{figures/invariance}
\smallskip
\noindent \textbf{Localization transfer.}
% \label{subsec:invariance}
An important benefit of formulating \ourmethod{} as a neural field optimization is the ability to trivially transfer localization results between different meshings. This ability is useful for many tasks in geometry processing which require an object to be re-triangulated, simplified, subdivided, or otherwise remeshed. Localization transfer is possible since our neural highlighter is represented as a field over the shape and is independent of any specific meshing. Although the neural highlighter is trained on mesh vertices, the resulting network encodes a smooth field and produces meaningful outputs for any
3D point on (or near) the mesh surface.

In \cref{fig:invariance}, we show an optimization of the \ourmethod{} on a single mesh triangulation (original) for the prompt `shoes'. We then apply the already-optimized neural highlighter to remeshed (middle) and subdivided (right) versions of the original mesh, showing the transferability of the selected region to different triangulations. This result demonstrates how \ourmethod{} is independent of the input mesh and that, once we have a localization for one mesh, we can trivially transfer it to any other meshing of the same object.

\smallskip
\noindent \textbf{Viewpoint robustness.}
Our method is robust to the primary view choice. This property is important for our localization task, as we may not know \textit{a priori} which view is ideal. In \cref{fig:viewpoints}, we perform our optimization using three different primary viewpoints: $0^\circ$, $90^\circ$, and $-90^\circ$ (viewpoints shown in blue). We then present predicted localizations, showing that for all three views, \ourmethod{} is able to accurately identify the target localization region, regardless of whether that region is visible from the primary view.

From the $-90^\circ$ primary view, the target region (the neck) is not visible. However, is is still visible with a low probability for views sampled from the Gaussian distribution around the primary view. This means that over the course of optimization, regions other than the neck are mostly seen while the target region is rarely visible. Nonetheless, our method manages to highlight the desired region, which implies its robustness to how frequently the target region for localization is seen. Furthermore, it shows that oversampling views where the target region is not visible does not negatively influence the optimization.

%%%%%%%%%%%%%%%%%%%%
%%% Applications %%%
%%%%%%%%%%%%%%%%%%%%
\subsection{Applications of \ourmethod{}}
\label{subsec:applications}

%\smallskip
\noindent \textbf{Selective editing.}
In \cref{fig:edit}, we show that it is possible to use \ourmethod{} to selectively edit a 3D object within a semantic region. This is applicable to techniques which incorporate global texture or material properties over the entire shape, such as in Text2Mesh~\cite{text2mesh} or MatCap~\cite{matcap}. Starting with different bare input meshes, we edit the entire shape using a global stylization technique~\cite{text2mesh}. Then, we use \ourmethod{} to select a text-specified region and incorporate the modifications \emph{only} in the selected area. Thus \ourmethod{} provides direct control over where to stylize shapes, enabling users to obtain localized stylizations based on semantic cues.

\smallskip
\noindent \textbf{Controlled stylization via composition.} 
Achieving compositionality with language models is a challenging task~\cite{dalle2}. For example, starting with a human mesh and using Text2Mesh~\cite{text2mesh} to stylize `\textit{Iron Man with the head of Steve Jobs and Yeti legs}', leads to muddled and undesirable results (\cref{fig:swap}, rightmost). Our method enables compositionality between different shape modifications by chaining simple concepts together (\cref{fig:swap}). Specifically, we decompose the desired modification into three separate attainable targets (`Iron Man', `Steve Jobs', and `Yeti'), which we stylize individually with Text2Mesh. We then utilize our \ourmethod{} to localize the text-specified regions. We achieve the desired composition by combining the highlighted regions together, obtaining clear boundaries between stylizations.

\smallskip
\noindent \textbf{Semantic segmentation.}
In \cref{fig:segmentation}, we show that our technique is not restricted to \emph{hallucinated highlighting} and is capable of localizing semantically-specified geometric regions. These text-driven localizations identify unique geometric parts \textit{without} utilizing any 3D datasets or part labels.

\input{figures/segmentation}

%%%%%%%%%%%%%%%%%%
%%% Components %%%
%%%%%%%%%%%%%%%%%%
\subsection{Components of \ourmethod{}}
\label{subsec:components}
\noindent \textbf{Ablation study.}
Several components are key for facilitating \ourmethod{}. We provide ablation results in \cref{fig:ablation} to demonstrate the effect of our design choices. First, using a direct optimization of the vertex color (\textit{direct}) instead of optimizing a neural field results in splotchy highlight artifacts. Since the neural field has a spectral bias towards smooth solutions~\cite{spectralbias}, omitting it leads to an undesired noisy output. Second, removing the probability weighted blending (\textit{no blend}) and instead coloring vertices using only two distinct values also produces a noisy highlight pattern. Without a continuous color blend, the gradients become ill-conditioned and unstable, leading to highlight artifacts and irregular localization boundaries. Lastly, similar to previous works~\cite{clipdraw, text2mesh}, we observe that without 2D perspective augmentations (\textit{no augs}), \ourmethod{} outputs degenerate solutions. The ablation study emphasizes the importance of our key design choices in \ourmethod{} for its ability to highlight a coherent and localized region on the input shape.

\input{figures/initialization}

\smallskip
\noindent \textbf{Prompt formulation and CLIP understanding.}
Our prompt formulation combined with our coloring scheme results in the correct association between objects and their properties, a known challenge when using CLIP~\cite{dalle2}. In \cref{fig:CLIP_scores}, we analyze the CLIP score for two different prompts: `\textit{gray chair with highlighted back}' (left) and `\textit{blue chair with red back}' (right). For each prompt, we  measure the CLIP similarity to renders of both the correct assignment and flipped assignment.

We observe that our prompt formulation (`\textit{gray chair with highlighted back}') results in a higher average CLIP score for the correct assignment. In contrast, when specifying colors in the prompt (`\textit{blue chair with red back}') and styling the mesh accordingly, we see \textit{higher} CLIP scores for the \textit{flipped} association. Using the same gray and yellow renders (left), we also compare to a prompt specifying colors (`gray chair with yellow back') and find that the higher CLIP score corresponds to the flipped selection (data not shown).

We also measure the CLIP scores for our standard prompt formulation: `gray chair with highlighted back', replacing the yellow color in the rendering with other colors, such as red and blue, and find that the correct selection has a higher CLIP score (data not shown). To conclude, our prompt formulation (i.e., the use of the term `highlighted') coincides with CLIP's understanding and \ourmethod{} is robust to the highlight color.

\input{figures/CLIP_scores}

\smallskip
\noindent \textbf{Network initialization.}
Initializing the network such that the object is partially highlighted (i.e., with highlight probability equal to $0.5$) is important for obtaining desirable results. In \cref{fig:initialization}, we show the optimization of our method for the target text prompt `belt' using three different initializations. Our method (middle) initializes all output probabilities near $0.5$ by random weight initialization of the network. We compare to initializing the output probabilities to $0$ (left) or $1$ (right), in which we set the weights of the last layer to $0$, and the bias to $0$ or $1$, respectively.

For the initialization to both $0.5$ and $1$, a highlight color is uniformly present on the styled mesh, whereas with $0$, the mesh is gray with no highlight. Consequently, we hypothesize that the presence of highlight color at initialization is important for CLIP's supervision.

%%%%%%%%%%%%%%%%%%%
%%% Limitations %%%
%%%%%%%%%%%%%%%%%%%
\subsection{Limitations}
\label{subsec:limitations}
\input{figures/prompt_robustness}

\ourmethod{} is robust to variations of the object specification in the target prompt. However, there should still be a logical connection between the 3D shape and its description. \cref{fig:prompt_robustness} shows results for a camel mesh and the target highlight `shinguards'. For each optimization, we use a slightly different target prompt by varying the \textit{object} specification. The prompts are of the form ``[\textit{object}] \textit{with highlighted shinguards}", where [\textit{object}] is replaced with \textit{camel}, \textit{pig}, \textit{animal}, or \textit{chair}.

In \cref{fig:prompt_robustness}, we observe that with object specifications that resemble the geometry of camel, such as pig and animal, \ourmethod{} accurately localizes the desired region. However, for a description that is incompatible with the object's geometry (i.e., referring to a camel as a chair), our method does not produce meaningful results. This result sheds light on \ourmethod{}'s robustness to text descriptions: \ourmethod{} is able to reason about a mesh even when its description is not perfectly accurate, provided that it is sufficiently similar to the true description (i.e., referring to a camel mesh as a pig).
\\

%% file: figures/ablation.tex
\begin{figure}[b]
    \centering
    \newcommand{\pl}{3.5}
    \newcommand{\plt}{-1}
    \begin{overpic}[width=\columnwidth]{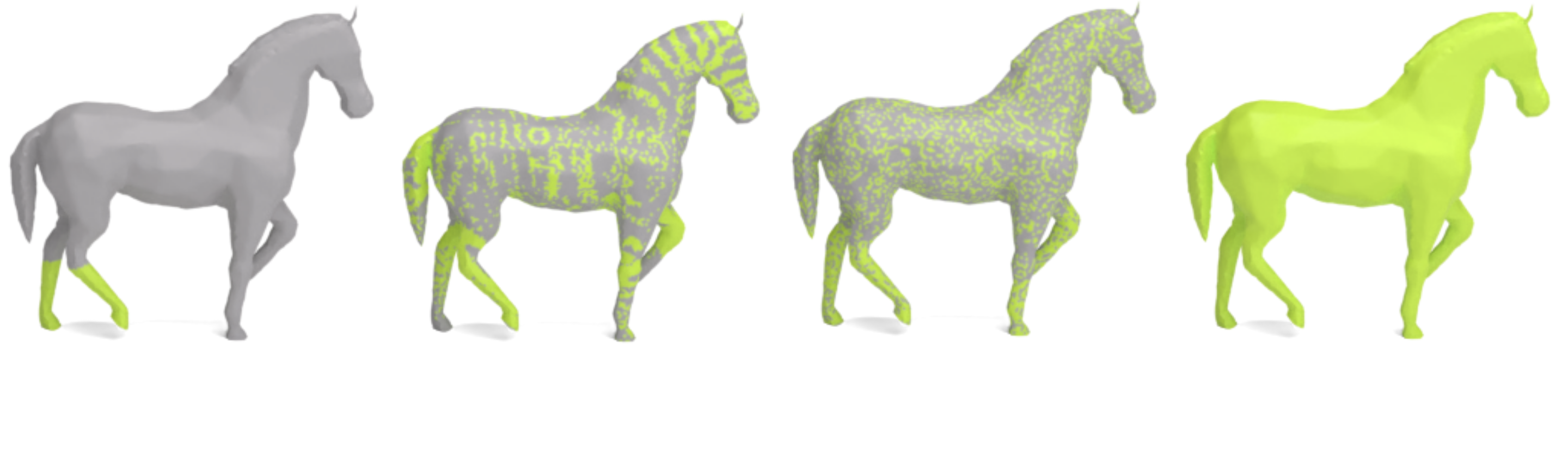}
    \put(7,  \pl){\textcolor{black}{\textit{\textbf{full}}}}
    \put(30,  \pl){\textcolor{black}{\textit{direct}}}
    \put(53,  \pl){\textcolor{black}{\textit{no blend}}}
    \put(80,  \pl){\textcolor{black}{\textit{no augs}}}
    \put(5,  \plt){\textcolor{black}{\textbf{0.332}}}
    \put(30,  \plt){\textcolor{black}{0.319}}
    \put(56,  \plt){\textcolor{black}{0.297}}
    \put(82,  \plt){\textcolor{black}{0.287}}
    \end{overpic}
    \caption{\textbf{Ablation experiments.}  We present ablation results for target text `shoes' using our system (\textit{\textbf{full}}), direct optimization (\textit{direct}), without probability-weighted blending (\textit{no blend}), and without 2D augmentations (\textit{no augs}). Resulting CLIP scores shown below each image.}
    \label{fig:ablation}
    \vspace{20pt}
\end{figure}

%% file: tables/clip_metrics.tex
\begin{table}[t!]
\small
 \begin{center}
    \begin{tabular}{l c c}
     \toprule
     CLIP Model & ViT-L/14 $\uparrow$ & ViT-B/16 $\uparrow$ \\
     \midrule
     LSeg~\cite{lseg} & 18.75 & 6.25\\
     Text2LIVE~\cite{text2live} & 43.75 & 31.25\\
     Ours & \textbf{81.25} & \textbf{43.75}\\
     \bottomrule
    \end{tabular}
    \caption{\textbf{Highlight fidelity.} We extend two image-based approaches LSeg~\cite{lseg} (segmentation) and Text2LIVE~\cite{text2live} (localized editing) to the highlighting task and report CLIP R-Precision.} \label{table:clip_metric}
 \end{center}
 
\end{table}

%% file: figures/swap.tex
\begin{figure}
    \centering
    \newcommand{\rowone}{-3}
    \begin{overpic}[width=\columnwidth]{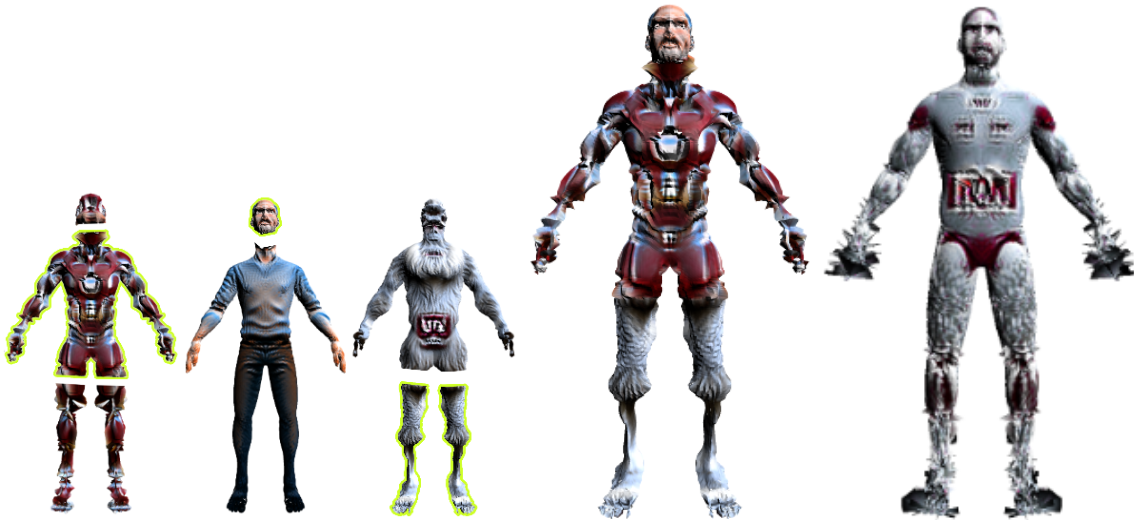}
    \put(3,  \rowone){\textit{Base}}
    \put(18,  \rowone){\textit{Head}}
    \put(35,  \rowone){\textit{Legs}}
    \put(55,  \rowone){Ours}
    \put(76,  \rowone){Text2Mesh}
    \end{overpic}
    \captionof{figure}{\textbf{Controlled stylization.} Given three different stylizations of the same object, we use \ourmethod{} to select different regions and combine them together (Ours). Attempting to achieve this composition with a holistic approach leads to an undesirable result (Text2Mesh~\cite{text2mesh}).}
    \label{fig:swap}
\end{figure}
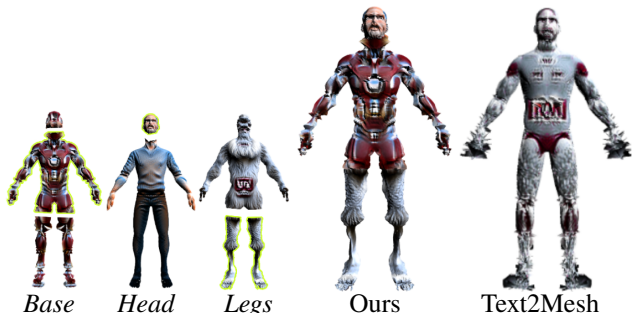

%% file: figures/invariance.tex
\begin{figure}[b]
    \centering
    \newcommand{\rowone}{-3}
    \begin{overpic}[width=\columnwidth]{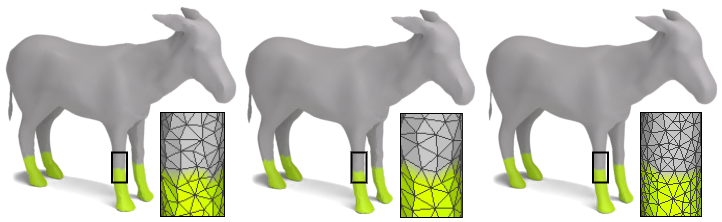}
    \put(7,  \rowone){Original}
    \put(40,  \rowone){Remeshed}
    \put(72,  \rowone){Subdivided}
    \end{overpic}
    \captionof{figure}{\textbf{Localization transfer.} We optimize our neural highlighter on one mesh (original) for the prompt `shoes'. Once optimized, the network weights transfer the localization to different meshings of the same object (remeshed and subdivided).}
    \label{fig:invariance}
\end{figure}

%% file: figures/segmentation.tex
\begin{figure}
    \centering
    \newcommand{\rowtwo}{-3}
    \begin{overpic}[width=0.9\columnwidth]{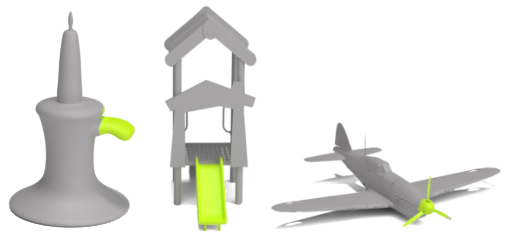}
    \put(10,  \rowtwo){Arm}
    \put(37,  \rowtwo){Slide}
    \put(70,  \rowtwo){Propeller}
    \end{overpic}
    \captionof{figure}{\textbf{Semantic Segmentation.} \ourmethod{} produces semantic segmentations for unique geometric parts \textit{without} any 3D dataset or annotations.}
    \label{fig:segmentation}
\end{figure}

%% file: figures/initialization.tex
\begin{figure}[b]
    \centering
    \newcommand{\rowone}{-4}
    \begin{overpic}[width=0.8\columnwidth]{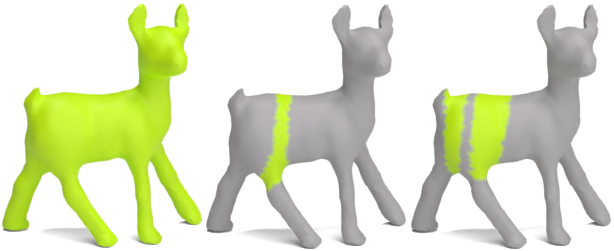}
    \put(17,  \rowone){$0$}
    \put(49,  \rowone){$0.5$}
    \put(84,  \rowone){$1$}
    \end{overpic}
    \captionof{figure}{\textbf{Network initialization}. We optimize \ourmethod{} for the text prompt `belt' using different initialization methods: using a default initialization where all output probabilities are near 0.5 (middle) or altering the final layer so that all outputs are $0$ (left) or $1$ (right). Initializing with $0$ or $1$ leads to an undesirable result.}
    \label{fig:initialization}
\end{figure}

%% file: figures/CLIP_scores.tex
\begin{figure}
    \centering
    \newcommand{\rowone}{-4}
    \newcommand{\rowtwo}{-9}
    \newcommand{\rowthree}{43}
    \begin{overpic}[width=0.8\columnwidth]{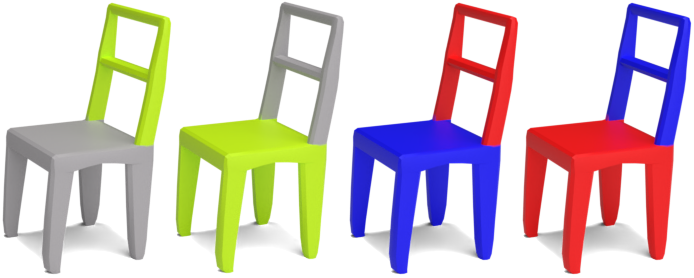}
    \put(6,  \rowone){Correct}
    \put(28,  \rowone){Flipped}
    \put(54,  \rowone){Correct}
    \put(82,  \rowone){Flipped}
    \put(7.5,  \rowtwo){\textbf{0.313}}
    \put(30,  \rowtwo){0.311}
    \put(55.5,  \rowtwo){0.315}
    \put(84,  \rowtwo){\textbf{0.318}}
    \put(12,  \rowthree){\textit{`\textit{highlighted back}'}}
    \put(70,  \rowthree){\textit{`\textit{red back}'}}
    \end{overpic}
    \vspace{12px}
    \captionof{figure}{\textbf{CLIP understanding.} We examine CLIP similarity scores for several prompt formulations targeting the `\textit{back}' of the chair while using the correct color assignment and where the coloring is flipped. For the prompt `\textit{gray chair with highlighted back}' (left) we observe that the CLIP score is higher for the \textbf{correct} assignment. For the the prompt `\textit{blue chair with red back}' (right) the CLIP score is higher for the \textbf{flipped} (incorrect) assignment.\dale{put somewhere so it is at top of page and overpic doesn't get crowded.}}
    \label{fig:CLIP_scores}
\end{figure}
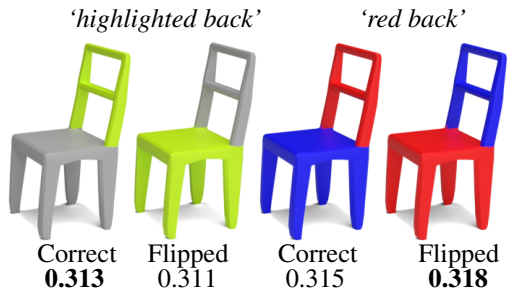

%% file: figures/prompt_robustness.tex
\begin{figure}[b]
    \centering
    \newcommand{\dblue}{\color[rgb]{0.56,0.82,0.92}}
    \newcommand{\red}{\color[rgb]{0.9,0.41,0.39}}
    \newcommand{\rowone}{-3}
    \newcommand{\rowtwo}{-7}
    \begin{overpic}[width=\columnwidth]{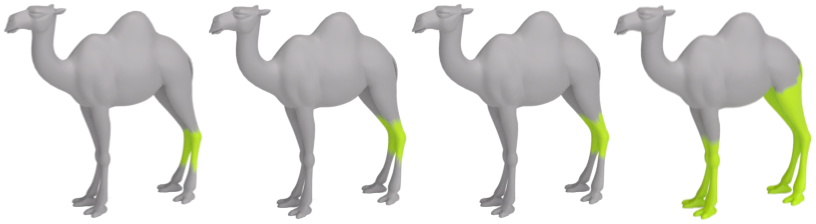}
    \put(6,  \rowone){Shinguards}
    \put(32,  \rowone){Shinguards}
    \put(56,  \rowone){Shinguards}
    \put(81,  \rowone){Shinguards}
    \put(10,  \rowtwo){\textbf{{\dblue{Camel}}}}
    \put(38,  \rowtwo){\textbf{{\dblue{Pig}}}}
    \put(60,  \rowtwo){\textbf{{\dblue{Animal}}}}
    \put(85,  \rowtwo){\textbf{{\red{Chair}}}}
    \end{overpic}
    \vspace{1px}
    \captionof{figure}{\textbf{Prompt generality}. Our system is robust to certain variations in {\dblue \textbf{object}} specifications. We achieve desirable results for the text input `\textit{{\dblue \textbf{camel}} with highlighted shinguards}' (left), as well as for other variations (`{\dblue \textbf{\textit{pig}}}' and `{\dblue \textbf{\textit{animal}}}'). If the object specification, such as `{\red \textbf{\textit{chair}}}', is incompatible with the input geometry, \ourmethod{} no longer produces meaningful results.}
    \label{fig:prompt_robustness}
\end{figure}
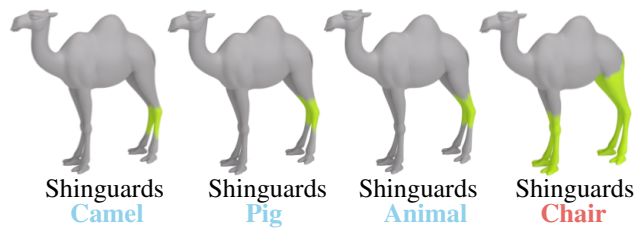

%% file: 05_conclusions.tex
\section{Conclusions}
\label{sec:conclusions}
We present a technique for \emph{highlighting} semantic regions on meshes using text as input, without any 3D datasets or 3D pre-training. \ourmethod{} can \textit{reason} about where to place a non-obviously related part on a 3D object (i.e. a hat on a candle). The ability to combine unconnected parts and objects together is reminiscent of ideas from \textit{image analogies}~\cite{liao2017visual, hertzmann2001image}. In this work, we show that we can identify part-concepts that are \textit{geometrically absent} from a shape, giving rise to our \emph{hallucinated highlighting} capability.

During neural optimization, our neural network infers a \textit{probability} which we use to blend the highlight color onto the mesh. The network-predicted probabilities are general, and provide a soft-segmentation which we show can be used for a variety of different applications (\cref{fig:swap,fig:edit}). In the future, we are interested in extending our framework to obtain part correspondence between shapes that differ topologically but are semantically related.

\section{Acknowledgments}

We thank the University of Chicago for providing the AI cluster resources, services, and the professional support of the technical staff. This work was also supported in part by gifts from Adobe Research. Finally, we would like to thank Richard Liu, Avery Zhou, and the members of 3DL for their thorough and insightful feedback on our work.

%% file: 07_supplementary.tex
%% Resets the figure and table to begin with A, e.g. Fig. A1
\newcommand{\beginsupplementary}{%
    \setcounter{section}{0}
	\renewcommand{\thesection}{A\arabic{section}}
	\renewcommand{\thesubsection}{\thesection.\arabic{subsection}}

	\renewcommand{\thetable}{A\arabic{table}}%
	\setcounter{table}{0}

	\renewcommand{\thefigure}{A\arabic{figure}}%
	\setcounter{figure}{0}
}

%\beginsupplementary
\appendix

\twocolumn[{%
 \centering
 {\bf \Large Supplementary Material for 3D Highlighter: Localizing Regions on 3D Shapes via Text Descriptions}
 \vspace{2em}
}]

%%%%%%%%% BODY TEXT
We provide additional information about our method. \cref{supp_additional_res} shows additional results and experiments we conducted with \ourmethod{}. \cref{supp_implementation} elaborates on our implementation details, including how we accomplish primary view selection, our network architecture, and our optimization scheme. \cref{supp_more_visual_results} shows highlights on additional mesh and prompt combinations.

\section{Additional Experiments and Details}
\label{supp_additional_res}

\noindent \textbf{Geometric edits.}
\ourmethod{} can be applied to create localized geometric edits by performing extrusion, stretching, deletion, and selection on localized regions (see ~\cref{fig:geometric_edits}). For extrusion, we scale vertices in the localized region along their normal vectors. With stretching, we shift the vertices in the localized region by some constant value. For deletion, we remove all vertices in the localized region as well as the faces they make up. For selection, we render only faces comprised entirely of vertices in the localized region. This application enables users to alter the geometry of semantic regions of 3D objects using only text.
\input{figures/geometric_edits}

\smallskip
\noindent \textbf{Multi-class semantic segmentation.}
Our method can be used to obtain multi-class semantic segmentations of 3D objects (see ~\cref{fig:graphcuts_segmentation}). This application takes advantage of \ourmethod{}'s ability to identify semantic regions. First we localize each semantic class on the object individually. Then we initialize a graph cut segmentation algorithm using our predicted probabilities for all classes. This algorithm outputs a segmentation of the vertices that is based on our predictions, but is smooth and conforms to the geometry of the mesh. This extension of \ourmethod{} allows users to acquire multi-class semantic segmentations for meshes with geometric parts not found in 3D datasets.
\input{figures/graphcuts_segmentation}

\input{figures/viewpoint_sampling}
\smallskip
\noindent \textbf{Viewpoint sampling.}
Our primary viewpoint sampling procedure is tailored to our specific localization task allowing it to produce more accurate highlights than other sampling methods. We evaluate three different viewpoint sampling procedures: ours (primary), anchor, and uniform sampling (\cref{fig:viewpoint_sampling}). Primary view sampling is described in \ref{subsec:guidance} in the main paper. 
Our primary view sampling is a variant of anchor view sampling (used in~\cite{text2mesh}), in which we sample all $n$ views from a Gaussian centered on the center view. The anchor sampling approach uses the center (anchor) view in each iteration along with $n-1$ Gaussian samples. Uniform sampling samples $n$ random views uniformly, independent of any center view. For the non-uniform approaches (primary and anchor) we evaluate each with two different center views, once where the target selection region is visible (blue) and once where it is not (orange).

We observe that our method produces similar results to the anchor method when using a center view where the text-specified target region is visible. However, when the target region is not visible, our method achieves more desirable results compared to anchor view sampling. This is because the anchor view sampling approach over-samples views near the anchor view and when this anchor view does not include the target region, it does not sample enough views of that region to effectively localize. Our method also produces better results than uniform sampling since uniform sampling results in many views from angles where the the shape is not recognizable to CLIP. By centering our sampling on a view where the shape is recognizable, we avoid impeding the optimization. Thus, our primary sampling approach strikes a balance in that we sample widely enough to sufficiently learn the target region while avoiding problematic views that are unrecognizable to CLIP and impede the optimization.
\input{figures/pe_ablation}

\smallskip
\noindent \textbf{Impact of positional encoding.}
As specified in \cref{subsec:neuralhighlighter} in the paper, we choose not to use a positional encoding. Although positional encoding has been shown to allow networks to learn high frequency features \cite{tancik2020fourfeat} and is commonly used in neural fields \cite{brownneuralfields}, our task actually benefits from low frequency predictions. In \cref{fig:pe_ablation}, we optimize \ourmethod{} on the target region `belt' both with and without a positional encoding. Using the positional encoding (right) gives the network too much freedom. This creates noisy highlight artifacts across the mesh. In extreme cases, the network can even hallucinate letters onto the mesh as seen in the figure. Without the positional encoding (left), the spectral bias results in a contiguous highlight region without any highlight artifacts.

\smallskip
\noindent \textbf{Quantitative evaluation details.}
In our quantitative evaluation, we measure the CLIP R-Precision of different localization methods on the highlighter task. Our CLIP R-Precision is defined as follows. We designate the set $T_p$ to be a set of $10$ possible target localizations. 
We also use 16 distinct meshes to create a dataset $D$ of $16$ different mesh/target localization pairs in which each mesh $M_i$ has a corresponding target localization $L_i$ where $L_i \in T_p$. To evaluate a method, we compute highlights for all pairs $(M_1, L_{i_1}), (M_2, L_{i_2}),...(M_{16}, L_{i_{16}})$ in $D$. Next, we use CLIP to attempt to retrieve the original target localization that was used to generate each highlight. To do so, we choose the target localization text in $T_p$ that has the highest CLIP similarity to the highlight. If this chosen target localization matches the target localization used to generate the highlight, then we consider that to be a successful retrieval. To obtain the CLIP R-Precision for a method, we report the percentage of mesh/target localization pairs in $D$ that were successfully retrieved.

As specified in \cref{subsec:fidelity}, we report the CLIP R-Precision for all methods using both the ViT-L/14 and ViT-B/16 CLIP models. Additionally, since our highlight is represented in 3D, we have to render our highlighted mesh into 2D. We do so using a different renderer than the one we use during optimization. By evaluating on different CLIP models and using a new renderer, we show that our highlight localizations are robust across different CLIP models and renderers.

As shown in \cref{table:clip_metric}, our method achieves higher CLIP R-Precision than both baselines. In addition to these quantitative results, \cref{fig:baselines} shows qualitative differences between the highlights of the different methods. From this figure, we can see that \ourmethod{} produces more accurate localizations than the baselines. Both Text2LIVE and LSeg also struggle to produce contiguous highlight regions. Additionally, LSeg frequently produces empty localization regions such as seen in the first example in the top left corner of \cref{fig:baselines}. The results of the baselines demonstrate the difficulty of the highlighter task.

\input{figures/baselines}

\input{figures/limitations.tex}

\smallskip
\noindent \textbf{CLIP understanding.}
\ourmethod{} relies on CLIP for supervision and thus is limited by biases in CLIP. There are cases where CLIP's understanding does not align with human visual understanding of where the region should be localized. In \cref{fig:limitations}, we use \ourmethod{} to localize the region `\textit{ears}' (left) on a bunny mesh. To a human observer, it is clear that the localization does not contain the bunny's ears: instead, the localization is a region on the side of the bunny's head. This is likely a result of CLIP more strongly associating ears with being placed on the side of a head than on top. In such cases, \ourmethod{} will provide poor localizations since it is based on CLIP's preferences. However, if we use \ourmethod{} to localize the region `\textit{headphones}' (right) on the same bunny mesh, we get a localization that has good visual correspondence to both `\textit{headphones}' and `\textit{ears}' (since the ideal localization for these two prompts on a bunny should look very similar). If we measure the CLIP similarity of both results to the text `\textit{gray bunny with highlighted ears}', we find that the `\textit{ears}' localization has higher CLIP similarity even though it has less visual correspondence. This explains why the `\textit{ears}' target region does not produce a localization like the one produced for the target region `\textit{headphones}'. Thus, when CLIP's biases lead to poor results on a given target region, we can often still obtain a good localization by running \ourmethod{} on a different specification of the target region.

\input{figures/consistency.tex}

\smallskip
\noindent \textbf{Optimization consistency and sensitivity.}
Depending on the combination of mesh and target localization region, \ourmethod{}'s optimization can vary in its sensitivity to non-determinism and thus its consistency (\cref{fig:consistency}). For some combinations of meshes and prompts, the supervision signal is very strong. As such, non-determinism has little to no impact and the optimization produces nearly identical results every run. However, for some mesh and prompt combinations, the supervision signal is weaker. As a result, the optimization is more sensitive to non-determinism and we see that the highlighted regions can differ significantly between runs.

\section{Implementation} \label{supp_implementation}

\input{figures/primary_view_selection}
\smallskip
\noindent \textbf{Choice of primary view.}
We use CLIP to automatically select our primary view (see ~\cref{fig:primary_view_sampling}). To sample our views during the rendering step, we need to choose a \textit{primary} view to center our view sampling distribution on. We want this view to correspond with CLIP's understanding of the underlying object as well as the target localization region. As such, we sample views uniformly around the object and for each rendered view, we encode it with CLIP and compare it to the encoded text target `A 3D render of a gray [object] with highlighted [target region]'. We then choose the view with the highest CLIP similarity to the text target to be our primary view.

\smallskip
\noindent \textbf{Neural highlighter architecture and implementation.}
Our neural highlighter is represented by an MLP with $6$ linear layers. The input dimension is $3$, encoding $(x,y,z)$ coordinates. Each linear layer has a width of $256$. After each of the first $5$ linear layers we apply a ReLU activation followed by a layer norm. After our $6$th and final linear layer, we instead apply a softmax activation that converts our output into a vector of probabilities. Thus, our final layer outputs an $n$ dimensional tensor where $n$ is the number of classes. Each element of the output tensor corresponds to the probability of the vertex belonging to that class. For the standard highlighter task, there are only $2$ classes: target region and not target region. Thus, there are 2 elements in the output vector and we can use the element of the output vector corresponding to the probability of belonging to the target region as the highlight probability described in the main paper.

\input{figures/viz-optim.tex}

\smallskip
\noindent \textbf{Optimization.}
We optimize the parameters of our neural highlighter using PyTorch's Adam optimizer with a constant learning rate of $1e^{-4}$. We train for 2500 iterations on a single A40 GPU which takes around 5 minutes to complete. See \cref{fig:viz-optim} for a visualization of the progression of predictions during the optimization process.

\section{More Visual Results}
\label{supp_more_visual_results}
We show highlights for additional combinations of meshes and prompts: \cref{fig:supp-animals} depicts highlights on animal meshes while \cref{fig:supp-objects} shows highlights on object meshes.
\input{figures/supp_animals.tex}
\input{figures/supp_objects.tex}

%%%%%%%%% REFERENCES
% {\small
% \bibliographystyle{ieee_fullname.bst}
% \bibliography{references.bib}
% }

% \end{document}

%% file: figures/geometric_edits.tex
\begin{figure}
    \centering
    \newcommand{\dblue}{\color[rgb]{0.56,0.82,0.92}}
    \newcommand{\dorange}{\color[rgb]{1,0.43,0.05}}
    \newcommand{\rowone}{50}
    \newcommand{\rowtwo}{-1}
    \begin{overpic}[width=\columnwidth]{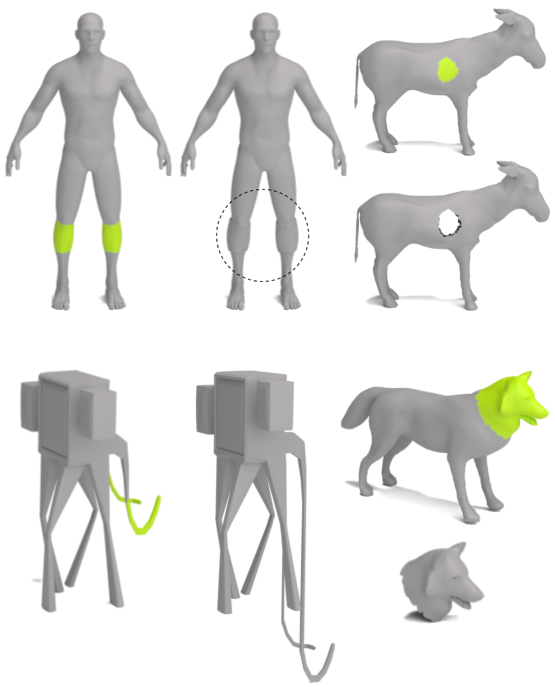}
    \put(12,  \rowone){Extruded \textit{Shinguards}}
    \put(52,  \rowone){Deleted \textit{Heart}}
    \put(13,  \rowtwo){Stretched \textit{Braids}}
    \put(53,  \rowtwo){Selected \textit{Head}}
    \end{overpic}
    \captionof{figure}{\textbf{Geometric edits}. Using regions selected with \ourmethod{},  we can perform localized geometric edits such as extrusion, stretching, deletion, and selection.}
    \label{fig:geometric_edits}
\end{figure}

%% file: figures/graphcuts_segmentation.tex
\begin{figure}
    \centering
    \newcommand{\yellow}{\color[rgb]{0.7, 0.85, 0.0}}
    \newcommand{\dorange}{\color[rgb]{1,0.43,0.05}}
    \newcommand{\rowone}{-3}
    \newcommand{\rowtwo}{-8}
    \begin{overpic}[width=\columnwidth]{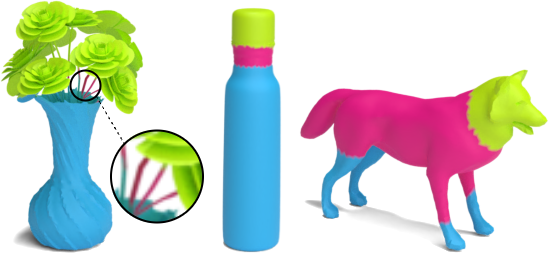}
    \put(0,  \rowone){Vase, Stems, Plants}
    \put(31,  \rowtwo){Body, Neck, Head}
    \put(60,  \rowone){Legs, Body, Head}
    \end{overpic}
    \vspace{1px}
    \captionof{figure}{\textbf{Multi-class semantic segmentation}. \ourmethod{} can be used to obtain semantic segmentations of 3D objects with unique geometric parts not found in any 3D dataset or annotations.}
    \label{fig:graphcuts_segmentation}
\end{figure}

%% file: figures/viewpoint_sampling.tex
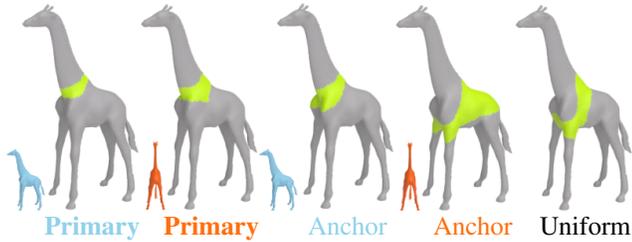
\begin{figure}
    \centering
    \newcommand{\dblue}{\color[rgb]{0.56,0.82,0.92}}
    \newcommand{\dorange}{\color[rgb]{1,0.43,0.05}}
    \newcommand{\rowone}{-3}
    \begin{overpic}[width=\columnwidth]{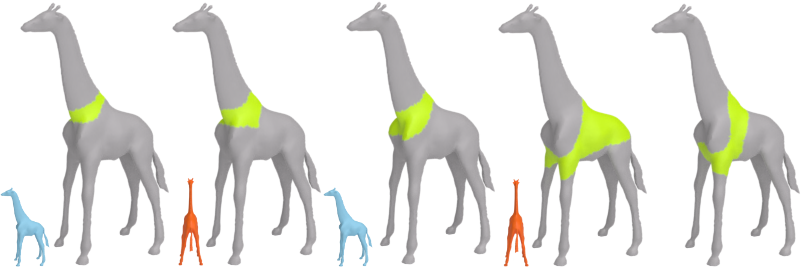}
    \put(6,  \rowone){{\dblue\textbf{Primary}}}
    \put(25,  \rowone){{\dorange\textbf{Primary}}}
    \put(48,  \rowone){{\dblue Anchor}}
    \put(68,  \rowone){{\dorange Anchor}}
    \put(85,  \rowone){Uniform}
    \end{overpic}
    \captionof{figure}{\textbf{Viewpoint sampling}. Results using different center view selection and sampling techniques for the target text: `necklace'. Our primary view sampling procedure produces better results than uniform sampling for center views where the selection region {\dblue \textbf{is}} (side on) and {\dorange \textbf{is not}} (facing away) visible. With a {\dblue \textbf{good center view}}, both our primary view method and the anchor view method produce accurate selections. However, with a {\dorange \textbf{bad center view}} our method produces more desirable results than the anchor view approach.}
    \label{fig:viewpoint_sampling}
\end{figure}

%% file: figures/pe_ablation.tex
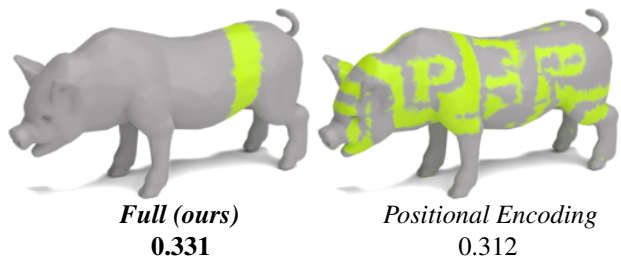
\begin{figure}[b]
    \centering
    \newcommand{\pl}{-3}
    \newcommand{\plt}{-8}
    \begin{overpic}[width=\columnwidth]{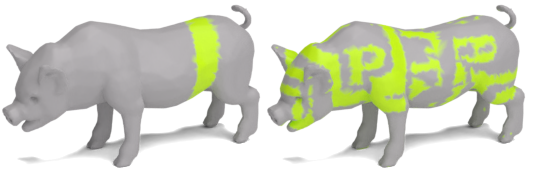}
    \put(18,  \pl){\textcolor{black}{\textit{\textbf{Full (ours)}}}}
    \put(60,  \pl){\textcolor{black}{\textit{Positional Encoding}}}
    \put(23,  \plt){\textcolor{black}{\textbf{0.331}}}
    \put(72,  \plt){\textcolor{black}{0.312}}
    \end{overpic}
    \vspace{5px}
    \caption{\textbf{Positional encoding impact}. We optimize \ourmethod{} for the target localization `\textit{belt}' with (\textit{Positional Encoding}) and without (\textit{\textbf{Standard}}) a positional encoding.}
    \label{fig:pe_ablation}
\end{figure}

%% file: figures/baselines.tex
\begin{figure}
    \centering
    \newcommand{\yellow}{\color[rgb]{0.7, 0.85, 0.0}}
    \newcommand{\dorange}{\color[rgb]{1,0.43,0.05}}
    \newcommand{\rowonea}{98}
    \newcommand{\rowoneb}{62}
    \newcommand{\rowonec}{26}
    \newcommand{\rowtwo}{68}
    \newcommand{\rowthree}{33}
    \newcommand{\rowfour}{-3}
    \begin{overpic}[width=\columnwidth]{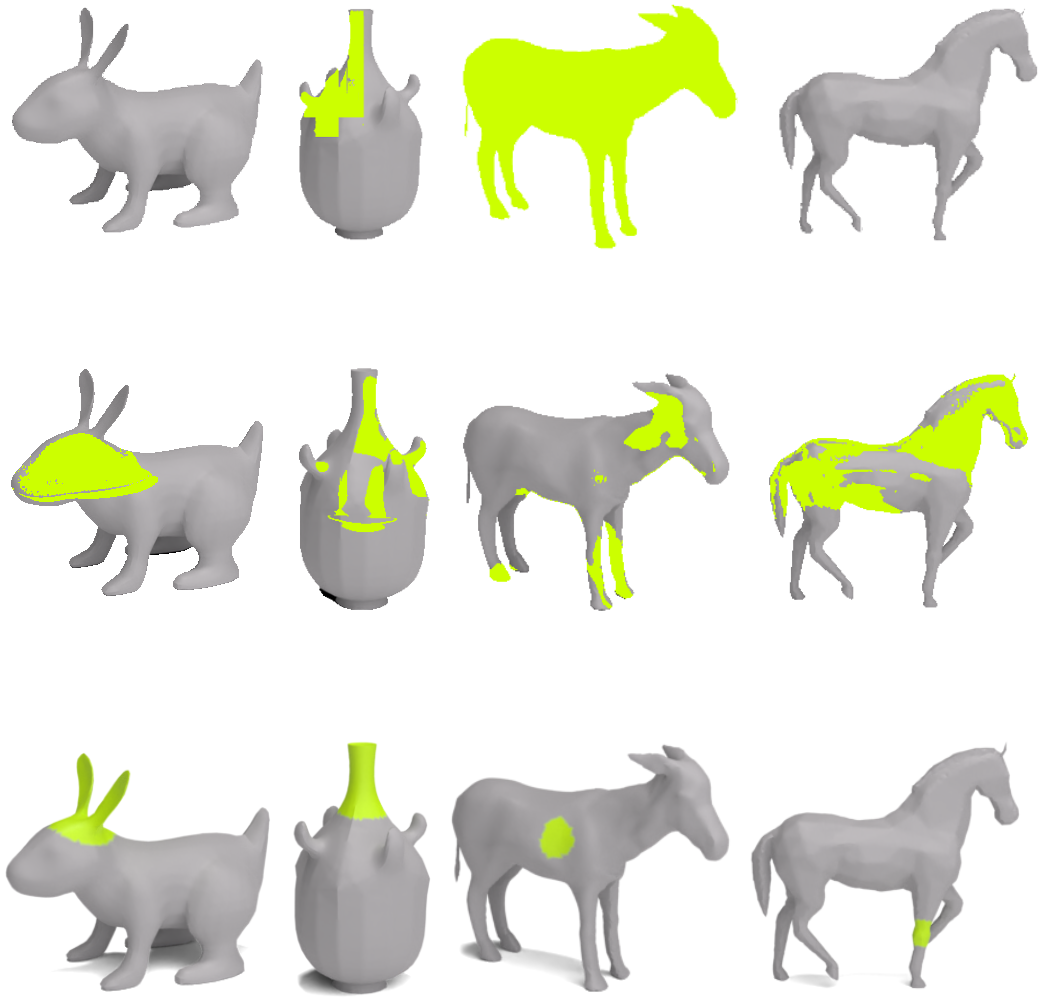}
    \put(0,  \rowonea){LSeg}
    \put(0,  \rowoneb){Text2LIVE}
    \put(0,  \rowonec){Ours}
    \put(12,  \rowtwo){Hat}
    \put(31,  \rowtwo){Hat}
    \put(50,  \rowtwo){Heart}
    \put(74,  \rowtwo){Wristwatch}
    \put(12,  \rowthree){Hat}
    \put(31,  \rowthree){Hat}
    \put(50,  \rowthree){Heart}
    \put(74,  \rowthree){Wristwatch}
    \put(12,  \rowfour){Hat}
    \put(31,  \rowfour){Hat}
    \put(50,  \rowfour){Heart}
    \put(74,  \rowfour){Wristwatch}
    \end{overpic}
    \vspace{1px}
    \captionof{figure}{\textbf{Qualitative Evaluation Examples}. Highlights on different meshes and prompts for LSeg~\cite{lseg}, Text2LIVE~\cite{text2live}, and \ourmethod{} (ours). Both baselines struggle on many mesh/prompt combinations. LSeg often outputs no selection region (as seen in the LSeg horse).}
    \label{fig:baselines}
\end{figure}
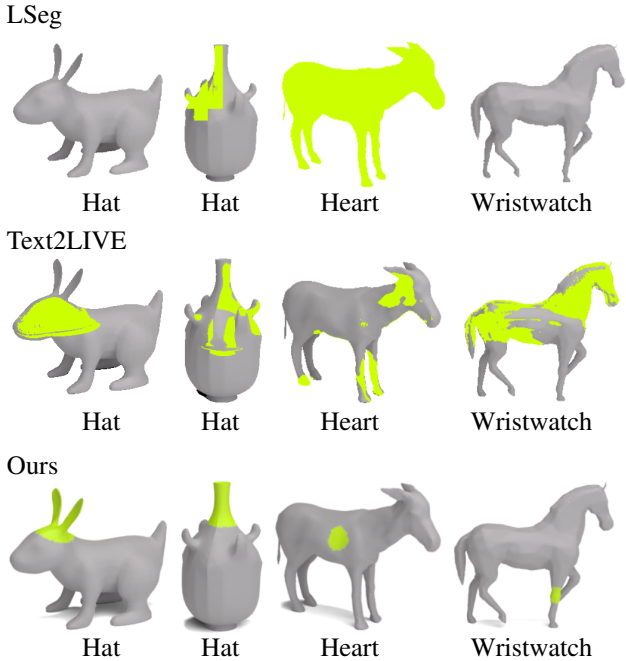

%% file: figures/limitations.tex
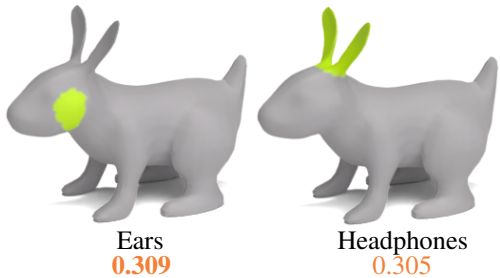
\begin{figure}
    \centering
    \newcommand{\rowone}{-4}
    \newcommand{\rowtwo}{-9}
    \begin{overpic}[width=0.8\columnwidth]{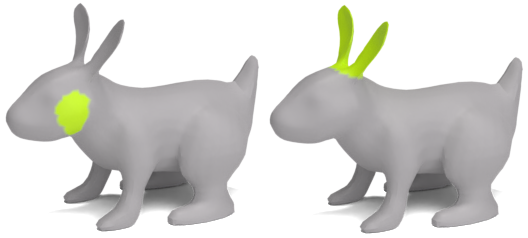}
    \put(23,  \rowone){Ears}
    \put(67,  \rowone){Headphones}
    \put(22,  \rowtwo){\textcolor{Orange}{\textbf{0.309}}}
    \put(74,  \rowtwo){\textcolor{Orange}{0.305}}
    \end{overpic}
    \vspace{12px}
    \captionof{figure}{\textbf{CLIP understanding}. The result of optimizing CLIP towards `\emph{headphones}' (right) results in a more `\emph{ear-like}' result compared to optimizing towards \emph{ears} (left). The \textcolor{Orange}{\textbf{CLIP similarity}} between the `\textit{ears}' text prompt and both highlighted meshes also confirms that CLIP's semantic association may not always correspond with the user's semantic association.}
    \label{fig:limitations}
\end{figure}

%% file: figures/consistency.tex
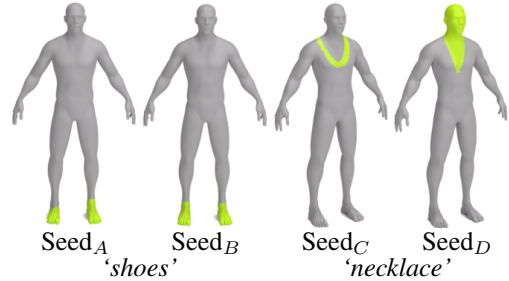
\begin{figure}
    \centering
    \newcommand{\rowone}{-4}
    \newcommand{\rowthree}{-9}
    \begin{overpic}[width=0.8\columnwidth]{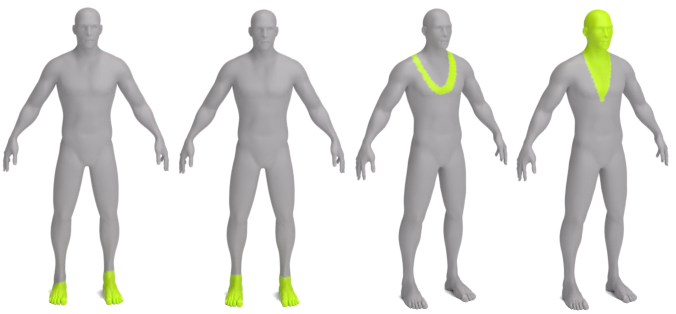}
    \put(6,  \rowone){Seed$_A$}
    \put(32,  \rowone){Seed$_B$}
    \put(58,  \rowone){Seed$_C$}
    \put(82,  \rowone){Seed$_D$}
    \put(18,  \rowthree){\textit{`\textit{shoes}'}}
    \put(66,  \rowthree){\textit{`\textit{necklace}'}}
    \end{overpic}
    \vspace{12px}
    \captionof{figure}{\textbf{Optimization sensitivity}. We observe that the results of \ourmethod{} are robust to different seeds when using certain target prompts (such as shoes, left); while other target prompts may produce more variable results (such as necklace, right).}
    \label{fig:consistency}
\end{figure}

%% file: figures/primary_view_selection.tex
\begin{figure}
    \centering
    \newcommand{\dblue}{\color[rgb]{0.56,0.82,0.92}}
    \newcommand{\dorange}{\color[rgb]{1,0.43,0.05}}
    \newcommand{\rowone}{-4}
    \begin{overpic}[width=\columnwidth]{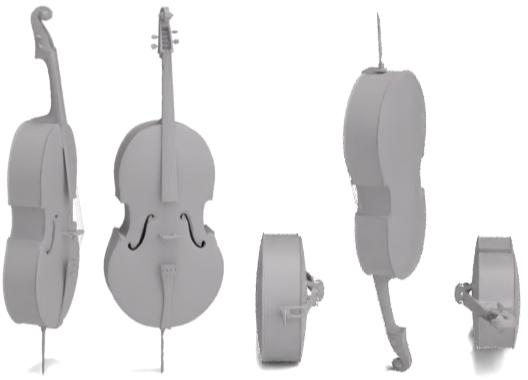}
    \put(5,  \rowone){0.34}
    \put(28,  \rowone){0.33}
    \put(50,  \rowone){0.24}
    \put(71,  \rowone){0.23}
    \put(90,  \rowone){0.18}
    \end{overpic}
    \vspace{1px}
    \captionof{figure}{\textbf{Automatic primary view selection}. We select our primary view by sampling views uniformly and choosing the view with the highest CLIP similarity to the text target. We show sampled views and their CLIP similarity score to the target prompt.}
    \label{fig:primary_view_sampling}
\end{figure}
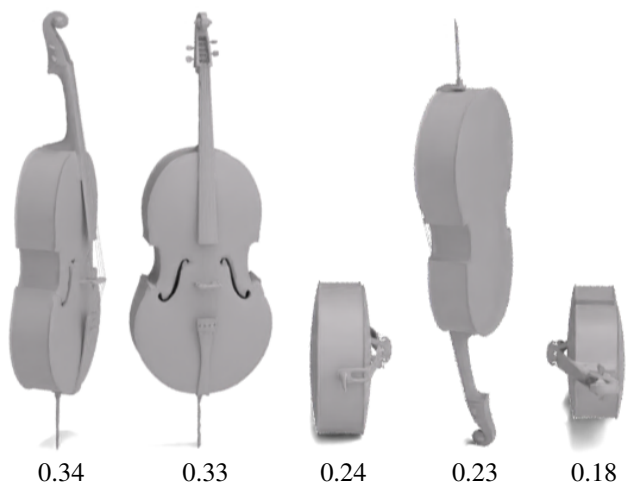

%% file: figures/viz-optim.tex
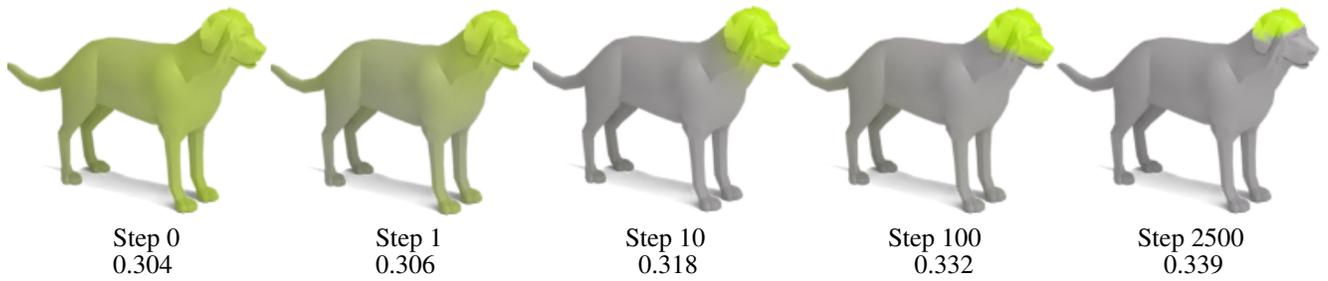
\begin{figure*}
    \centering
    \newcommand{\pl}{-2}
    \newcommand{\plt}{-4}
    \begin{overpic}[width=\textwidth]{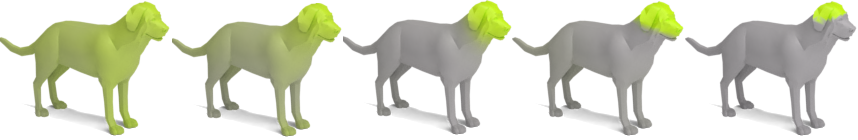}
    \put(8,  \pl){Step 0}
    \put(28,  \pl){Step 1}
    \put(47,  \pl){Step 10}
    \put(67,  \pl){Step 100}
    \put(86,  \pl){Step 2500}
    \put(8,  \plt){0.304}
    \put(28,  \plt){0.306}
    \put(48,  \plt){0.318}
    \put(69,  \plt){0.332}
    \put(88,  \plt){0.339}
    \end{overpic}
    \vspace{6px}
    \caption{\textbf{Optimization visualization}. We optimize \ourmethod{} on a mesh of a dog with the target localization `\textit{hat}' and visualize the predicted highlighted region at five steps throughout the optimization. We also report the CLIP similarity score at each step.}
    \label{fig:viz-optim}
\end{figure*}

%% file: figures/supp_animals.tex
\begin{figure*}[t!]
    \centering
    \newcommand{\rowone}{63}
    \newcommand{\rowtwo}{47}
    \newcommand{\rowthree}{32}
    \newcommand{\rowfour}{15}
    \newcommand{\rowfive}{-2}
    \begin{overpic}[width=\textwidth]{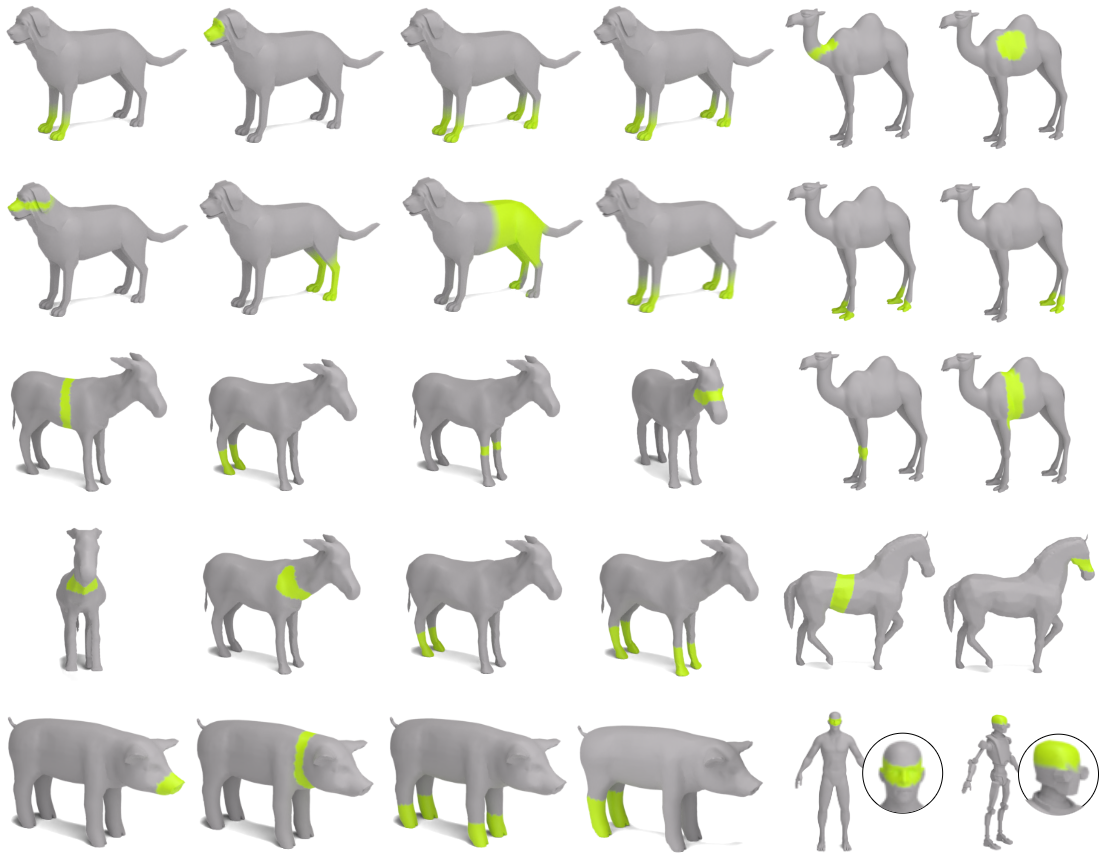}
    \put(3,  \rowone){\textcolor{black}{Wristwatch}}
    \put(24,  \rowone){\textcolor{black}{Mask}}
    \put(42,  \rowone){\textcolor{black}{Shoes}}
    \put(60,  \rowone){\textcolor{black}{Floor}}
    \put(76,  \rowone){\textcolor{black}{Necklace}}
    \put(91,  \rowone){\textcolor{black}{Heart}}
    \put(5,  \rowtwo){\textcolor{black}{Glasses}}
    \put(20,  \rowtwo){\textcolor{black}{Shinguards}}
    \put(42,  \rowtwo){\textcolor{black}{Cape}}
    \put(59,  \rowtwo){\textcolor{black}{Wheels}}
    \put(76,  \rowtwo){\textcolor{black}{Wheels}}
    \put(90,  \rowtwo){\textcolor{black}{Shoes}}
    \put(6,  \rowthree){\textcolor{black}{Belt}}
    \put(20,  \rowthree){\textcolor{black}{Shinguards}}
    \put(39,  \rowthree){\textcolor{black}{Wristwatch}}
    \put(58,  \rowthree){\textcolor{black}{Glasses}}
    \put(75,  \rowthree){\textcolor{black}{Wristwatch}}
    \put(91,  \rowthree){\textcolor{black}{Belt}}
    \put(4,  \rowfour){\textcolor{black}{Bowtie}}
    \put(20,  \rowfour){\textcolor{black}{Necklace}}
    \put(40,  \rowfour){\textcolor{black}{Wheels}}
    \put(58,  \rowfour){\textcolor{black}{Shoes}}
    \put(75,  \rowfour){\textcolor{black}{Saddle}}
    \put(88,  \rowfour){\textcolor{black}{Glasses}}
    \put(6,  \rowfive){\textcolor{black}{Mask}}
    \put(20,  \rowfive){\textcolor{black}{Necklace}}
    \put(40,  \rowfive){\textcolor{black}{Shoes}}
    \put(56,  \rowfive){\textcolor{black}{Shinguards}}
    \put(74,  \rowfive){\textcolor{black}{Eyeglasses}}
    \put(89,  \rowfive){\textcolor{black}{Hat}}
    \end{overpic}
    \vspace{1px}
    \caption{\textbf{Animal gallery}. Example highlights on meshes of dogs, goats, camels, horses, pigs, humans, and robots.}
    \label{fig:supp-animals}
\end{figure*}
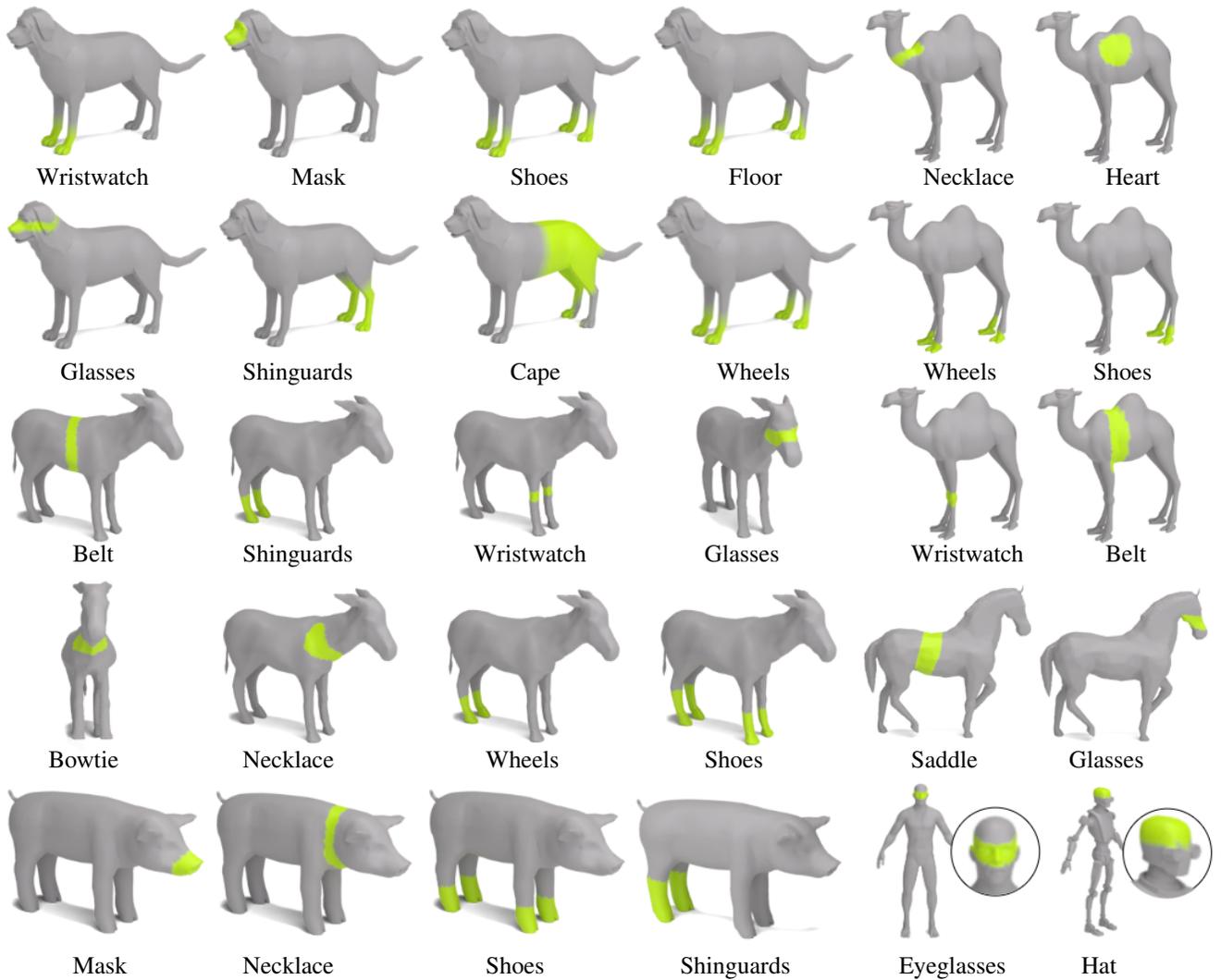

%% file: figures/supp_objects.tex
\begin{figure*}
    \centering
    \newcommand{\rowone}{39}
    \newcommand{\rowtwo}{0}
    \begin{overpic}[width=\textwidth]{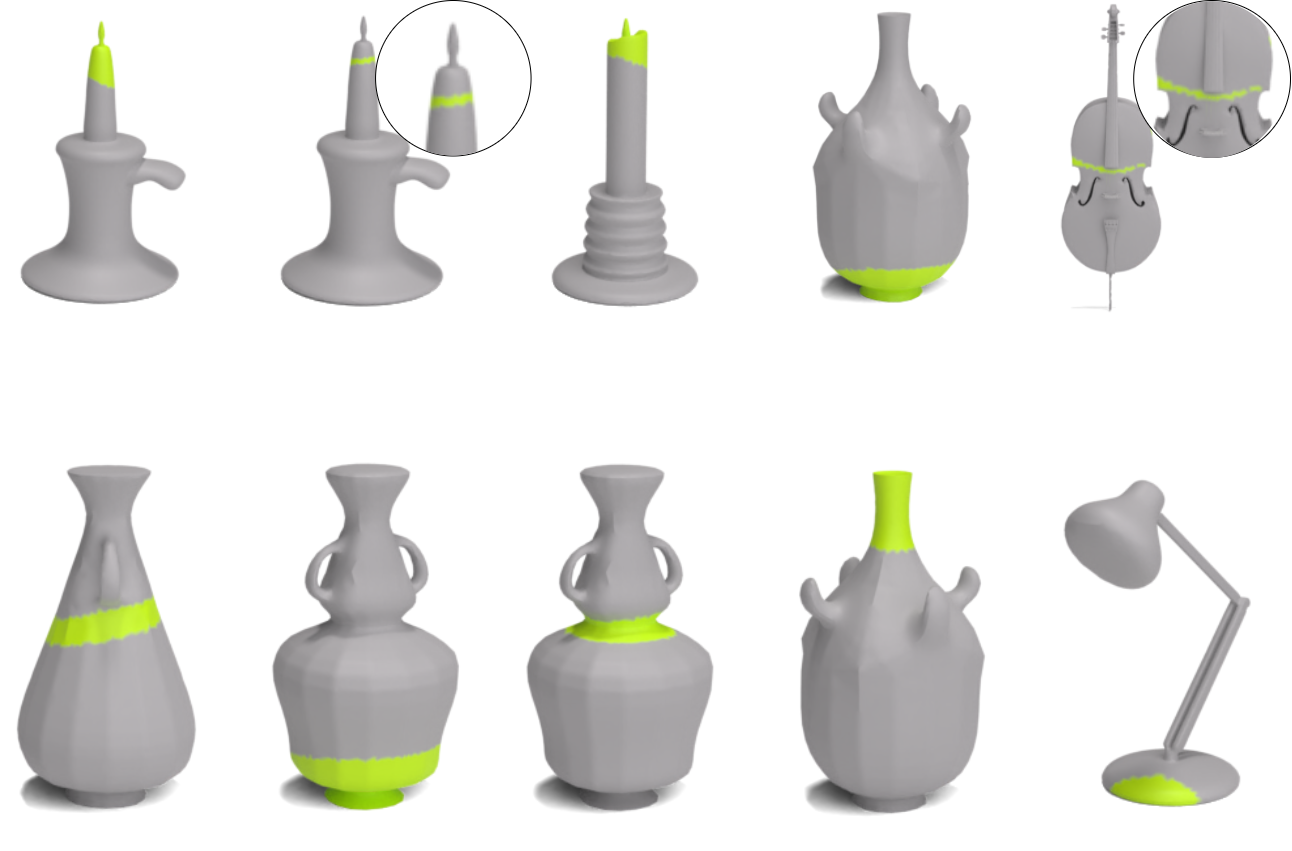}
    \put(6,  \rowone){\textcolor{black}{Hair}}
    \put(27,  \rowone){\textcolor{black}{Belt}}
    \put(47,  \rowone){\textcolor{black}{Hair}}
    \put(66,  \rowone){\textcolor{black}{Wheels}}
    \put(85,  \rowone){\textcolor{black}{Belt}}
    \put(6,  \rowtwo){\textcolor{black}{Belt}}
    \put(25,  \rowtwo){\textcolor{black}{Wheels}}
    \put(44,  \rowtwo){\textcolor{black}{Necklace}}
    \put(67,  \rowtwo){\textcolor{black}{Hair}}
    \put(87,  \rowtwo){\textcolor{black}{Shoes}}
    \end{overpic}
    \caption{\textbf{Object gallery}. Example highlights on meshes of candles, vases, instruments, and lamps.}
    \label{fig:supp-objects}
\end{figure*}

%% file: main.bbl
\begin{thebibliography}{10}\itemsep=-1pt

\bibitem{asafi2013weak}
Shmuel Asafi, Avi Goren, and Daniel Cohen-Or.
\newblock Weak convex decomposition by lines-of-sight.
\newblock {\em Computer graphics forum}, 32(5):23--31, 2013.

\bibitem{text2live}
Omer Bar-Tal, Dolev Ofri-Amar, Rafail Fridman, Yoni Kasten, and Tali Dekel.
\newblock Text2live: Text-driven layered image and video editing.
\newblock {\em arXiv preprint arXiv:2204.02491}, 2022.

\bibitem{chang2015shapenet}
Angel~X Chang, Thomas Funkhouser, Leonidas Guibas, Pat Hanrahan, Qixing Huang,
  Zimo Li, Silvio Savarese, Manolis Savva, Shuran Song, Hao Su, et~al.
\newblock Shapenet: An information-rich 3d model repository.
\newblock {\em arXiv preprint arXiv:1512.03012}, 2015.

\bibitem{chen2019learning}
Wenzheng Chen, Huan Ling, Jun Gao, Edward Smith, Jaakko Lehtinen, Alec
  Jacobson, and Sanja Fidler.
\newblock Learning to predict 3d objects with an interpolation-based
  differentiable renderer.
\newblock {\em Advances in Neural Information Processing Systems}, 32, 2019.

\bibitem{chen2019bae}
Zhiqin Chen, Kangxue Yin, Matthew Fisher, Siddhartha Chaudhuri, and Hao Zhang.
\newblock Bae-net: Branched autoencoder for shape co-segmentation.
\newblock In {\em Proceedings of the IEEE/CVF International Conference on
  Computer Vision}, pages 8490--8499, 2019.

\bibitem{cornea2007curve}
Nicu~D Cornea, Deborah Silver, and Patrick Min.
\newblock Curve-skeleton properties, applications, and algorithms.
\newblock {\em IEEE Transactions on visualization and computer graphics},
  13(3):530, 2007.

\bibitem{deng2020cvxnet}
Boyang Deng, Kyle Genova, Soroosh Yazdani, Sofien Bouaziz, Geoffrey Hinton, and
  Andrea Tagliasacchi.
\newblock Cvxnet: Learnable convex decomposition.
\newblock In {\em Proceedings of the IEEE/CVF Conference on Computer Vision and
  Pattern Recognition}, pages 31--44, 2020.

\bibitem{dey2004approximating}
Tamal~K Dey and Wulue Zhao.
\newblock Approximating the medial axis from the voronoi diagram with a
  convergence guarantee.
\newblock {\em Algorithmica}, 38(1):179--200, 2004.

\bibitem{clipdraw}
Kevin Frans, Lisa~B. Soros, and Olaf Witkowski.
\newblock Clipdraw: Exploring text-to-drawing synthesis through language-image
  encoders.
\newblock {\em ArXiv}, abs/2106.14843, 2021.

\bibitem{hanocka2019meshcnn}
Rana Hanocka, Amir Hertz, Noa Fish, Raja Giryes, Shachar Fleishman, and Daniel
  Cohen-Or.
\newblock Meshcnn: A network with an edge.
\newblock {\em ACM Transactions on Graphics (TOG)}, 38(4):90:1--90:12, 2019.

\bibitem{hertzmann2001image}
Aaron Hertzmann, Charles~E Jacobs, Nuria Oliver, Brian Curless, and David~H
  Salesin.
\newblock Image analogies.
\newblock In {\em Proceedings of the 28th annual conference on Computer
  graphics and interactive techniques}, pages 327--340, 2001.

\bibitem{hoffman1984parts}
Donald~D Hoffman and Whitman~A Richards.
\newblock Parts of recognition.
\newblock {\em Cognition}, 18(1-3):65--96, 1984.

\bibitem{hong2022threedcg}
Yining Hong, Yilun Du, Chunru Lin, Josh Tenenbaum, and Chuang Gan.
\newblock 3d concept grounding on neural fields.
\newblock In {\em Annual Conference on Neural Information Processing Systems},
  2022.

\bibitem{subdivnet}
Shi{-}Min Hu, Zheng{-}Ning Liu, Meng{-}Hao Guo, Junxiong Cai, Jiahui Huang,
  Tai{-}Jiang Mu, and Ralph~R. Martin.
\newblock Subdivision-based mesh convolution networks.
\newblock {\em {ACM} Trans. Graph.}, 41(3):25:1--25:16, 2022.

\bibitem{dreamfields}
Ajay Jain, Ben Mildenhall, Jonathan~T Barron, Pieter Abbeel, and Ben Poole.
\newblock Zero-shot text-guided object generation with dream fields.
\newblock In {\em Proceedings of the IEEE/CVF Conference on Computer Vision and
  Pattern Recognition}, pages 867--876, 2022.

\bibitem{kaick2014shape}
Oliver~Van Kaick, Noa Fish, Yanir Kleiman, Shmuel Asafi, and Daniel Cohen-Or.
\newblock Shape segmentation by approximate convexity analysis.
\newblock {\em ACM Transactions on Graphics (TOG)}, 34(1):1--11, 2014.

\bibitem{clipmesh}
Nasir~Mohammad Khalid, Tianhao Xie, Eugene Belilovsky, and Popa Tiberiu.
\newblock Clip-mesh: Generating textured meshes from text using pretrained
  image-text models.
\newblock {\em SIGGRAPH Asia 2022 Conference Papers}, December 2022.

\bibitem{kobayashi2022distilledfeaturefields}
Sosuke Kobayashi, Eiichi Matsumoto, and Vincent Sitzmann.
\newblock Decomposing nerf for editing via feature field distillation.
\newblock {\em arXiv}, 2022.

\bibitem{lahav2020meshwalker}
Alon Lahav and Ayellet Tal.
\newblock Meshwalker: Deep mesh understanding by random walks.
\newblock {\em ACM Transactions on Graphics (TOG)}, 39(6):1--13, 2020.

\bibitem{lseg}
Boyi Li, Kilian~Q Weinberger, Serge Belongie, Vladlen Koltun, and Rene Ranftl.
\newblock Language-driven semantic segmentation.
\newblock In {\em International Conference on Learning Representations}, 2022.

\bibitem{liao2017visual}
Jing Liao, Yuan Yao, Lu Yuan, Gang Hua, and Sing~Bing Kang.
\newblock Visual attribute transfer through deep image analogy.
\newblock {\em arXiv preprint arXiv:1705.01088}, 2017.

\bibitem{lien2007approximate}
Jyh-Ming Lien and Nancy~M Amato.
\newblock Approximate convex decomposition of polyhedra.
\newblock In {\em Proceedings of the 2007 ACM symposium on Solid and physical
  modeling}, pages 121--131, 2007.

\bibitem{maron2017convolutional}
Haggai Maron, Meirav Galun, Noam Aigerman, Miri Trope, Nadav Dym, Ersin Yumer,
  Vladimir~G Kim, and Yaron Lipman.
\newblock Convolutional neural networks on surfaces via seamless toric covers.
\newblock {\em ACM Trans. Graph.}, 36(4):71--1, 2017.

\bibitem{text2mesh}
Oscar Michel, Roi Bar-On, Richard Liu, Sagie Benaim, and Rana Hanocka.
\newblock Text2mesh: Text-driven neural stylization for meshes.
\newblock In {\em Proceedings of the IEEE/CVF Conference on Computer Vision and
  Pattern Recognition}, pages 13492--13502, 2022.

\bibitem{milano2020primal}
Francesco Milano, Antonio Loquercio, Antoni Rosinol, Davide Scaramuzza, and
  Luca Carlone.
\newblock Primal-dual mesh convolutional neural networks.
\newblock {\em Advances in Neural Information Processing Systems}, 33:952--963,
  2020.

\bibitem{kaichun2018partnet}
Kaichun Mo, Shilin Zhu, Angel~X. Chang, Li Yi, Subarna Tripathi, Leonidas~J.
  Guibas, and Hao Su.
\newblock Partnet: A large-scale benchmark for fine-grained and hierarchical
  part-level 3d object understanding, 2018.

\bibitem{park2021benchmark}
Dong~Huk Park, Samaneh Azadi, Xihui Liu, Trevor Darrell, and Anna Rohrbach.
\newblock Benchmark for compositional text-to-image synthesis.
\newblock In {\em Thirty-fifth Conference on Neural Information Processing
  Systems Datasets and Benchmarks Track (Round 1)}, 2021.

\bibitem{pytorch}
Adam Paszke, Sam Gross, Soumith Chintala, Gregory Chanan, Edward Yang, Zachary
  DeVito, Zeming Lin, Alban Desmaison, Luca Antiga, and Adam Lerer.
\newblock Automatic differentiation in pytorch.
\newblock In {\em NIPS-W}, 2017.

\bibitem{dreamfusion}
Ben Poole, Ajay Jain, Jonathan~T. Barron, and Ben Mildenhall.
\newblock Dreamfusion: Text-to-3d using 2d diffusion.
\newblock {\em arXiv}, 2022.

\bibitem{CLIP}
Alec Radford, Jong~Wook Kim, Chris Hallacy, Aditya Ramesh, Gabriel Goh,
  Sandhini Agarwal, Girish Sastry, Amanda Askell, Pamela Mishkin, Jack Clark,
  et~al.
\newblock Learning transferable visual models from natural language
  supervision.
\newblock {\em arXiv preprint arXiv:2103.00020}, 2021.

\bibitem{spectralbias}
Nasim Rahaman, Aristide Baratin, Devansh Arpit, Felix Draxler, Min Lin, Fred
  Hamprecht, Yoshua Bengio, and Aaron Courville.
\newblock On the spectral bias of neural networks.
\newblock In Kamalika Chaudhuri and Ruslan Salakhutdinov, editors, {\em
  Proceedings of the 36th International Conference on Machine Learning},
  volume~97 of {\em Proceedings of Machine Learning Research}, pages
  5301--5310. PMLR, 09--15 Jun 2019.

\bibitem{dalle2}
Aditya Ramesh, Prafulla Dhariwal, Alex Nichol, Casey Chu, and Mark Chen.
\newblock Hierarchical text-conditional image generation with clip latents.
\newblock {\em arXiv preprint arXiv:2204.06125}, 2022.

\bibitem{Toys4k}
James~Matthew Rehg.
\newblock Toys4k 3d object dataset, 2022.
\newblock https://github.com/rehg-lab/lowshot-shapebias/tree/main/toys4k.

\bibitem{shamir2008survey}
Ariel Shamir.
\newblock A survey on mesh segmentation techniques.
\newblock {\em Computer graphics forum}, 27(6):1539--1556, 2008.

\bibitem{sharp2022diffusionnet}
Nicholas Sharp, Souhaib Attaiki, Keenan Crane, and Maks Ovsjanikov.
\newblock Diffusionnet: Discretization agnostic learning on surfaces.
\newblock {\em ACM Transactions on Graphics (TOG)}, 41(3):1--16, 2022.

\bibitem{canonical_capsules}
Weiwei Sun, Andrea Tagliasacchi, Boyang Deng, Sara Sabour, Soroosh Yazdani,
  Geoffrey~E Hinton, and Kwang~Moo Yi.
\newblock Canonical capsules: Self-supervised capsules in canonical pose.
\newblock In M. Ranzato, A. Beygelzimer, Y. Dauphin, P.S. Liang, and J.~Wortman
  Vaughan, editors, {\em Advances in Neural Information Processing Systems},
  volume~34, pages 24993--25005. Curran Associates, Inc., 2021.

\bibitem{tancik2020fourfeat}
Matthew Tancik, Pratul~P. Srinivasan, Ben Mildenhall, Sara Fridovich-Keil,
  Nithin Raghavan, Utkarsh Singhal, Ravi Ramamoorthi, Jonathan~T. Barron, and
  Ren Ng.
\newblock Fourier features let networks learn high frequency functions in low
  dimensional domains.
\newblock {\em NeurIPS}, 2020.

\bibitem{matcap}
Hideki Todo, Ken Anjyo, and Shun'Ichi Yokoyama.
\newblock Lit-sphere extension for artistic rendering.
\newblock {\em Vis. Comput.}, 29(6–8):473–480, jun 2013.

\bibitem{turbosquid}
TurboSquid.
\newblock Turbosquid 3d model repository, 2021.
\newblock https://www.turbosquid.com/.

\bibitem{coseg_2011}
Oliver van Kaick, Andrea Tagliasacchi, Oana Sidi, Hao Zhang, Daniel Cohen-Or,
  Lior Wolf, and Ghassan Hamarneh.
\newblock Prior knowledge for part correspondence.
\newblock {\em Computer Graphics Forum}, 30(2):553–562, 2011.

\bibitem{modelnet}
Zhirong Wu, Shuran Song, Aditya Khosla, Fisher Yu, Linguang Zhang, Xiaoou Tang,
  and Jianxiong Xiao.
\newblock 3d shapenets: A deep representation for volumetric shapes.
\newblock In {\em Proceedings of the IEEE conference on computer vision and
  pattern recognition}, pages 1912--1920, 2015.

\bibitem{brownneuralfields}
Yiheng Xie, Towaki Takikawa, Shunsuke Saito, Or Litany, Shiqin Yan, Numair
  Khan, Federico Tombari, James Tompkin, Vincent Sitzmann, and Srinath Sridhar.
\newblock Neural fields in visual computing and beyond.
\newblock {\em Computer Graphics Forum}, 2022.

\bibitem{yi2017syncspeccnn}
Li Yi, Hao Su, Xingwen Guo, and Leonidas~J Guibas.
\newblock Syncspeccnn: Synchronized spectral cnn for 3d shape segmentation.
\newblock In {\em Proceedings of the IEEE conference on computer vision and
  pattern recognition}, pages 2282--2290, 2017.

\bibitem{yu2019partnet}
Fenggen Yu, Kun Liu, Yan Zhang, Chenyang Zhu, and Kai Xu.
\newblock Partnet: A recursive part decomposition network for fine-grained and
  hierarchical shape segmentation.
\newblock In {\em Proceedings of the IEEE/CVF Conference on Computer Vision and
  Pattern Recognition}, pages 9491--9500, 2019.

\bibitem{skeleton_intrinsic}
Qian Zheng, Zhuming Hao, Hui Huang, Kai Xu, Hao Zhang, Daniel Cohen-Or, and
  Baoquan Chen.
\newblock Skeleton-intrinsic symmetrization of shapes.
\newblock {\em Computer Graphics Forum}, 34(2):275--286, 2015.

\bibitem{Thingi10K}
Qingnan Zhou and Alec Jacobson.
\newblock Thingi10k: A dataset of 10,000 3d-printing models.
\newblock {\em arXiv preprint arXiv:1605.04797}, 2016.

\bibitem{zhu2020adacoseg}
Chenyang Zhu, Kai Xu, Siddhartha Chaudhuri, Li Yi, Leonidas~J Guibas, and Hao
  Zhang.
\newblock Adacoseg: Adaptive shape co-segmentation with group consistency loss.
\newblock In {\em Proceedings of the IEEE/CVF Conference on Computer Vision and
  Pattern Recognition}, pages 8543--8552, 2020.

\end{thebibliography}
